\documentclass[pdflatex,sn-mathphys-num]{sn-jnl}% Math and Physical Sciences Numbered Reference Style
\usepackage{graphicx}%
\usepackage{multirow}%
\usepackage{amsmath,amssymb,amsfonts}%
\usepackage{amsthm}%
\usepackage{mathrsfs}%
\usepackage[title]{appendix}%
\usepackage{xcolor}%
\usepackage{textcomp}%
\usepackage{manyfoot}%
\usepackage{booktabs}%
\usepackage{algorithm}%
\usepackage{algorithmicx}%
\usepackage{algpseudocode}%
\usepackage{listings}%
\theoremstyle{thmstyleone}%
\theoremstyle{thmstyletwo}%

\theoremstyle{thmstylethree}%

\begin{document}

\title[Article Title]{X-ray and neural network based in-situ identification of the melt pool during the additive manufacturing of a stainless steel part}


\author*[1,2]{\fnm{Loïc} \sur{Jegou}}\email{loic.jegou@esrf.fr}

\author[2]{\fnm{Valérie} \sur{Kaftandjian}}\email{valerie.kaftandjian@insa-lyon.fr}

\author[1]{\fnm{Thomas} \sur{Elguedj}}\email{thomas.elguedj@insa-lyon.fr}
\author[2]{\fnm{Mohamed} \sur{Tahraoui}}\email{mohamed.tahraoui@insa-lyon.fr}
\author[2]{\fnm{Philippe} \sur{Duvauchelle}}\email{philippe.duvauchelle@insa-lyon.fr}
\author[1]{\fnm{Mady} \sur{Guillemot}}\email{mady.guillemot@insa-lyon.fr}

\affil*[1]{\orgdiv{Univ. Lyon}, \orgname{INSA Lyon, CNRS, Lamcos, UMR5259}, \orgaddress{\city{Villeurbanne}, \postcode{69621}, \country{France}}}

\affil[2]{\orgdiv{INSA Lyon}, \orgname{LVA, UR677}, \orgaddress{\city{Villeurbanne}, \postcode{69621}, \country{France}}}

%\affil[3]{\orgdiv{Department}, \orgname{Organization}, \orgaddress{\street{Street}, \city{City}, \postcode{610101}, \state{State}, \country{Country}}}


\abstract{Laser Metal Deposition with Powder (LMDp) is an additive manufacturing technique used for repairing metal components or producing parts with intricate geometries. However, a comprehensive understanding of the melt pool dynamics, which significantly influences the final properties of LMDp-fabricated parts, remains limited. Non-destructive testing is highly valuable for conducting in-situ controls during manufacturing. X-ray imaging offers the ability to penetrate metallic parts and detect defects such as porosity. In the context of additive manufacturing, X-rays can be employed to visualize the shape of the melt pool during the fabrication process. The contrast between the liquid and solid phases, due to their density differences, should be observable in the radioscopy images.
The experimental setup required to perform such a test on an industrial additive manufacturing installation consists of a movable X-ray source that produces polychromatic beams, a detector, and extensive lead shielding to ensure X-ray safety. In-situ observations of the melt pool were conducted during the deposition of ten successive layers of stainless steel 316L (SS316L). The polychromatic nature of the X-ray beam, however, rendered traditional image analysis methods ineffective for detecting contrast variations. To address this challenge, neural networks trained on simulated data (thermal and X-ray) were employed, providing a solution to identify the melt pool in low-contrast radioscopic images. The architecture inspired by VGG16 demonstrated promising results, confirming the potential for in-situ non-destructive testing using X-ray imaging in industrial additive manufacturing processes.}

\keywords{Additive manufacturing, Direct Energy Deposition, Stainless steel, X-ray, Melt pool, Convolutional neural networks}



\maketitle

\section{Introduction}\label{chap:introduction}

%================================================================
%Additive manufacturing
Additive manufacturing (AM) is defined by the American Society for Testing and Materials as: "the process of joining materials to make objects from 3D model data, usually layer upon layer, as opposed to subtractive manufacturing methodologies, such as traditional machining". While this definition applies to all types of materials, this work specifically focuses on metallic parts. Regarded as the new industrial revolution \cite{Berman2012}, AM has continuously evolved to address current industrial demands, such as rapid production, material efficiency, and the ability to create complex, custom-designed components. Today, a lot of different technologies exist \cite{Frazier2014}, and this article focuses on Laser Metal Deposition with Powder \cite{Ahn2021} (LMDp). 
LMDp involves using a highly focused laser to melt a localized region of a substrate, while continuously delivering powdered feedstock material into the melt pool \cite{Svetlizky2021a}. During this process, there is a complex interaction between the powder, laser, and melt pool which is strongly influenced by the operating parameters (such as powder flowrate, laser power and scanning speed). These parameters significantly affect the quality of the final part in terms of geometry, residual stress \cite{Lu2019}, and metallurgical properties \cite{Marya2018}.
%================================================================
%Melt pool
The melt pool serves as a reliable indicator of defect formation \cite{Scime2019} and overall quality. Monitoring it in real-time during the manufacturing process is highly valuable, both for predicting defect locations (such as porosity, as demonstrated by \citet{Khanzadeh2019} using infrared images) and for adjusting operating parameters during the fabrication, as seen in studies using infrared sensors \cite{Gibson2020b} or CMOS cameras \cite{Akbari2019}. 

Numerous imaging techniques have been developed to monitor the melt pool's morphology \cite{Sun2020, Ding2016, Smurov2013}, its temperature \cite{Liu2014, Everton2016}, or both simultaneously \cite{Jegou2023}. While these methods provide critical insights into the melt pool's behavior, they are limited to surface observations, leaving the subsurface dynamics unobserved. This is where X-rays step in, offering the ability to peer through the material and capture details beneath the surface.
%================================================================
%Xray in the industry
In the industry, X-rays are extensively used to perform Non-Destructive Testing (NDT) \cite{hanke2008} to inspect manufactured parts for defects, such as porosity. Radioscopy is an imaging technique that creates two-dimensional grayscale images that reflect the degree to which X-rays interact with an object \cite{Letang}. This method examines contrasts resulting from variations in thickness, chemical composition, or density. In metal AM, the idea is to analyze the melt pool by leveraging the contrast created by the density differences between the liquid and solid phases. Prior to its application in AM, X-rays were used to study melt pools in welding and laser cutting processes with similar objectives \cite{Abt2011, Boley2013}.  While porosities were easily visible in these  studies, the melt pool contours were often indistinct due to low contrast between the two phases.
\citet{Yamada2012} addressed this issue using synchrotron radiation, where X-ray beams are far brighter (at least a million times more photons) than those produced by portable X-ray sources. This led to numerous studies of LMDp conducted at synchrotron facilities, such as the Advanced Photon Source in the USA \cite{ Wolff2019, Wolff2022, lindenmeyer2021template, webster2023pore} and Diamond Light Source in England \cite{Chen2021, Chen2021_1, Chen2021_2}. These studies provided valuable insights  into porosity formation mechanisms,  powder particle interactions above the melt pool, particle trajectories influenced by laser-induced pressure gradients, and melt pool dynamics. However, synchrotrons are huge installations that generate intense X-ray beams with the capability to filter them into monochromatic rays. The downside is that AM setups must be miniaturized to fit within these facilities \cite{dass2022laser}, which results in systems with characteristics that differ from those used in industrial environments.
The primary goal of this study is to make X-ray NDT feasible on an industrial additive manufacturing setup, using a movable X-ray source. 

Portable X-ray sources, unlike synchrotrons, produce polychromatic beams that are significantly less bright, resulting in radioscopic images with lower contrast. The next challenge is to develop a method for identifying the melt pool in these low-contrasted grayscale images. This issue closely parallels challenges in the medical field, where tumors or pathological features must be detected in X-ray scans with similarly low contrast.
%================================================================
%Xray in the medical 
Several image segmentation techniques exist, including thresholding \cite{Sahoo1988}, region-based methods\cite{Karthick2014}, boundary-based approaches\cite{Zhu2007}, and hybrid techniques \cite{Nyma2012}. Currently, automatic segmentation methods are favored due to their lack of reliance on manual adjustments. Otsu's thresholding, known for its efficiency, has inspired many researchers \cite{Ng2006, Fan2012}. However, the effectiveness of these methods can be compromised by the poor quality of the radioscopy, the size of the tumor or the level of contrast \cite{Swetha2016}. Artificial Intelligence (AI) provides new approaches to deal with these segmentation challenges, achieving higher success rates \cite{Chowdhary2020, Liu2019}. Typically, computer vision problems are solved with Convolutional Neural Networks (CNN). A comprehensive review of these architectures is provided by \citet{Zhang2017}. Among these, auto-encoders (a specific type of CNN) have gained popularity in recent years for image processing tasks \cite{tschannen2018}. All these techniques inspired the approach described in this paper for identifying the melt pool in low-contrast radioscopies.
Training such models requires a substantial amount of data, currently lacking in LMDp. To address this limitation, a simplified thermal simulation of the process has been used to generate a model of a typical melt pool, followed by a simulation of an X-ray shot onto the melt pool. 

Finally, different architectures of \textit{auto-encoders have been trained on simulations of in-situ radioscopies} of the LMDp process, and subsequently applied to \textit{experimental radioscopy to identify the melt pool} during the deposition of ten successive layers of stainless steel parts.
The paper begins with a brief overview of X-ray theory, followed by a detailed description of the complete experimental setup, including the specific safety measures related to X-ray use. After calibrating the system, the first image acquisition of in-situ radioscopies of the melt pool in LMDp is conducted. Various image analysis techniques are employed to identify the melt pool but prove to be ineffective. In the next phase, a full dataset of simulated radioscopies of the melt pool is generated to train three auto-encoders inspired by the architectures of LeNET5, AlexNET, and VGG16. Upon completion of the training, these auto-encoders successfully identify the melt pool in simulated radioscopies, regardless of its size or position within the image. Finally, the same auto-encoders are directly applied to the experimental images without additional training, yielding promising results from the model inspired by VGG16. 

\section{Methodology}\label{chap:methodology}
\subsection{Theory}

\label{susec:theory}
\subsubsection{X-ray formation}
\begin{figure}[h]
	\begin{center}
	\includegraphics[width=0.75\textwidth]{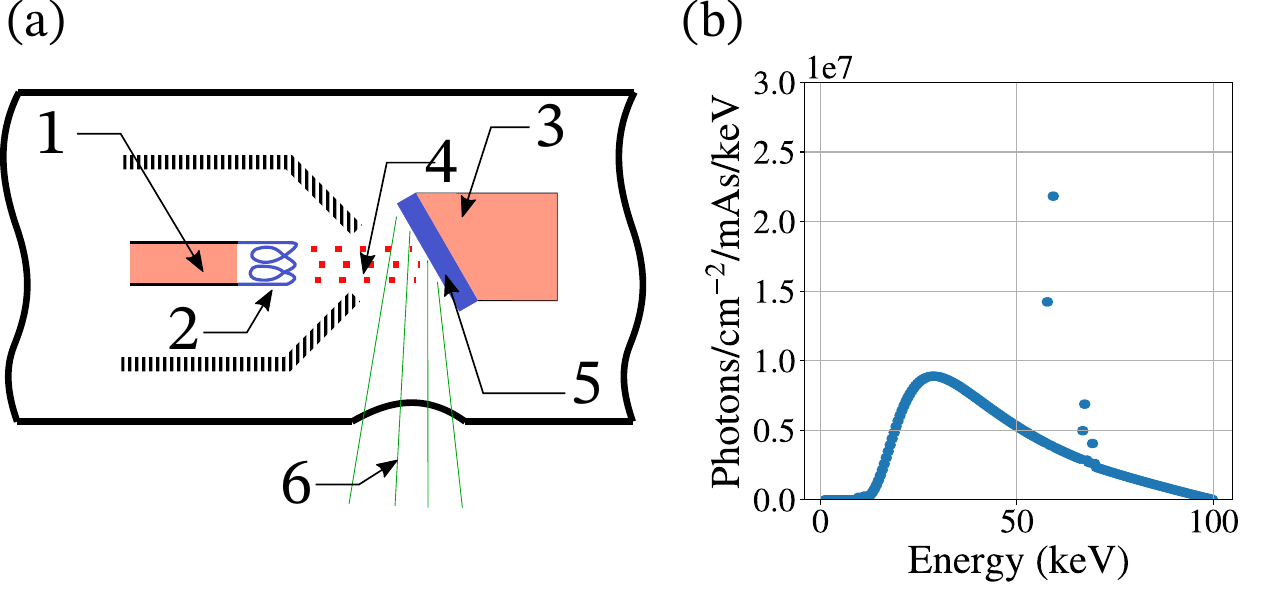}
	\end{center}
	\caption{\label{fig:source} (a) Schematic of an x-ray source. (b) The energy spectrum of an incident X-ray beam with a high voltage of 100 kV.}
\end{figure}

X-rays are high-energy photons characterized by wavelengths shorter than $10^{-8}$ m. In both medical and industrial applications, the most prevalent method for generating X-rays is through the use of portable X-ray generators (figure~\ref{fig:source}~(a)).

Electrons move freely in a closed circuit looping through the anode and the tungsten filament (indicated as (1) and (2) on figure~\ref{fig:source}). When a voltage difference is applied between the cathode and the anode (3), electrons are emitted (4) and collide with the tungsten target (5). This interaction generates the primary X-ray beam (6), which can subsequently be filtered and collimated according to specific requirements. 
The photons in the primary beam exhibit a broad range of energy levels, with the maximum energy determined by the applied voltage difference (figure~\ref{fig:source}~(b)).
The variation in photon energy implies that they will interact differently with objects.

\subsubsection{X-ray interactions and attenuation}
At industrial voltage levels (approximately 100 kV), the predominant interaction between photons and matter is attributed to the photoelectric effect. 
On a broader scale, X-ray interactions with matter can be viewed as a statistical problem: what is the probability that an X-ray will interact with electrons over a specified length of a material? Beer-Lambert's law provides a simplified answer to this question. For a homogeneous material with density $\rho$ (g.cm$^{-3}$), an attenuation coefficient $\mu$ (cm$^{-1}$), and  a given thickness $x$ (cm), the intensity of the transmitted X-ray beam $I_1$ (for a given energy E) is determined as follows (equation~\ref{eq:beer_lambert}).

\begin{equation}\label{eq:beer_lambert}
	I_1(E) = I_0(E) \times exp(-\frac{\mu}{\rho}(E) \times \rho x).
\end{equation}

$\frac{\mu}{\rho}$ is the massic attenuation coefficient (in cm$^2$.g$^{-1}$) and is known for every element of the periodic table for a given energy \cite{HubbellJohnHandSeltzer1995}.

This equation is valid for monochromatic X-ray beams (where all photons possess the same energy E). Such beams are typically only produced in synchrotrons. In this study, the X-ray beam is polychromatic, meaning that the photons from the incident beam will not undergo uniform attenuation, resulting in noisier and less contrasted images.
Setting aside this complexity for the moment, for a given material placed in an X-ray beam, the measured signal on the detector will solely depend on the thickness and density of that material. For a fixed thickness, even a slight variation in density will lead to a small change in the signal detected in the radioscopy. This low contrast becomes more challenging to detect with polychromatic beams, as the transmitted signal $I_1$ is noisier. Nonetheless, the objective of this paper is to leverage this characteristic to identify the melt pool during the additive manufacturing process.

\section{Experimental setup}\label{subsec:setup}

\subsection{X-ray source and detector}\label{sec:source}
\begin{figure}[h]
	\begin{center}
	\includegraphics[width=0.75\textwidth]{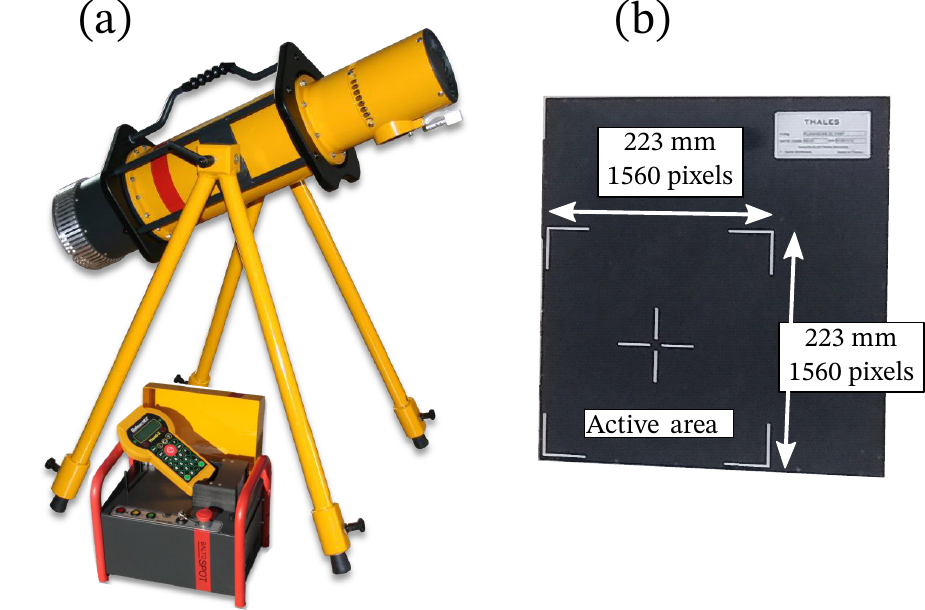}
	\end{center}	\caption{\label{fig:source_detector} (a) X-ray source. (b) X-ray flatpanel detector (\textit{the pictures are not scaled}).}
\end{figure}

The X-ray source utilized in this study is a portable generator (LLX200-DA-1) from \textit{Balteau NDT} (figure~\ref{fig:source_detector}~(a)). This source is commonly employed in open environments, such as construction sites, for the examination of industrial components. It is capable of producing X-rays with a voltage ranging from 5 to 200 kV and a current intensity of 0.5 to 12 mA, with a maximum power output of 900 W (additional specifications are available on the \href{https://www.balteau-ndt.com/baltospot/?product=llx200-da}{manufacturer's website}). X-ray emission is initiated using a wireless remote control. The detector employed in this experiment is a flat panel (flashscan 23 from \textit{Thales}) featuring a wide active area that enables a broad field of view (figure~\ref{fig:source_detector}~(b)). Its specifications are detailed in the subsequent table (table~\ref{tab:det_char}).

\begin{table}[h]
\caption{Characteristics of the detector.}\label{tab:det_char}%
\begin{tabular}{@{}ll@{}}
\toprule
Characteristics & Value\\
\midrule
Dimensions of the active area (pixels)    & 1560 $\times$ 1560 \\
Size of a pixel ($\mu$m)    & 143 $\times$ 143   \\
Scintillator   & Gd$_2$O$_2$S (Gadox)   \\
Voltage (kV) & 40 to 150 \\
Year of manufacture & 2006 \\
\botrule
\end{tabular}
%\footnotetext{Source: This is an example of table footnote. This is an example of table footnote.}
%\footnotetext[1]{Example for a first table footnote. This is an example of table footnote.}
%\footnotetext[2]{Example for a second table footnote. This is an example of table footnote.}
\end{table}

%    \begin{table}[pos = H]
%      \centering
%      \begin{tabular}[x]{|m{11em}||m{8em}|}
%      \hline
%      \textbf{Characteristic} & \textbf{Value}\\
%      \hline
%      Dimensions of the active area (pixels) & 1560 $\times$ 1560 \\
%      \hline
%      Size of a pixel ($\mu$m) & 143 $\times$ 143 \\
%      \hline
%      Scintillator & Gd$_2$O$_2$S (Gadox) \\
%      \hline
%      Voltage (kV) & 40 to 150 \\
%      \hline
%	  Acquisition rate (Hz) & 0.1 \\
%      \hline
%      Year of manufacture & 2006 \\
%      \hline
%      \end{tabular}
%      \caption{Characteristics of the detector.}
%      \label{tab:det_char}
%    \end{table}
    
\subsection{Additive manufacturing setup}\label{sususec:LASCOL}
The additive manufacturing setup is located in an open area of 3 square meters. It features a 6-axis robot arm (Staubli RX160) equipped with a set of outer and inner nozzles (Precitec - ZM YC50 DAS 0.5 II and ZM YC50 DIS II). The laser and the powder are routed to the nozzles via a diode laser (Laserline LMD2000) with a wavelength ranging from 900 to 1200 nm and a beam diameter of 1.4 mm, as well as a powder feeder (Oerlikon Twin 150) that contains two distinct tanks for different materials. The powder used in this study is stainless steel (SS316L), with particles of size \(106 \pm 45 \mu m\). The evolution of the density of SS316L as a function of temperature is discussed in detail in the appendix \ref{sususec:density}.
There is one single feeding nozzle, which is coaxial to the laser. The focal distance of the laser is slightly superior to the focal distance of the powder 

\begin{figure}[h]
	\begin{center}
	\includegraphics[width=0.75\textwidth]{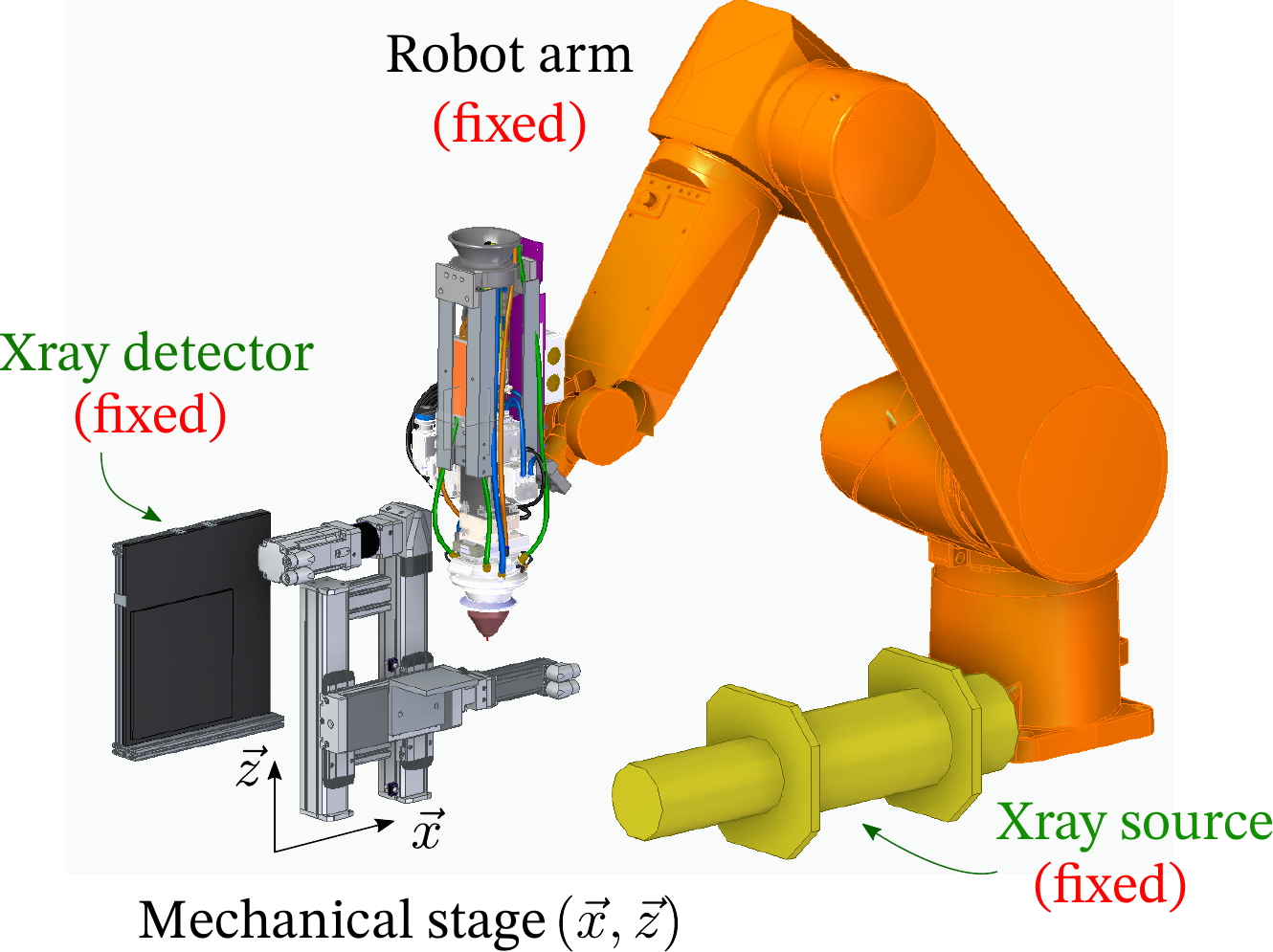}
	\end{center}	\caption{\label{fig:robot_source_det} 3D visualization of the robot arm and the x-ray equipment.}
\end{figure}

The X-ray source and detector are positioned on either side of the robotic arm (figure~\ref{fig:robot_source_det}). In this configuration, the laser remains stationary while a mechanical stage beneath the substrate moves horizontally (along $\vec{x}$) and vertically (along $\vec{z}$) to deposit each layer onto the previous one. The \textit{laser being fixed} also ensures that the \textit{melt pool remains stationary}, allowing for extended X-ray exposure times (\textit{i.e.} lower intensities from the X-ray source) for detailed studies. 

\subsection{Safety measures}
\begin{figure}[h]
	\begin{center}
	\includegraphics[width=0.75\textwidth]{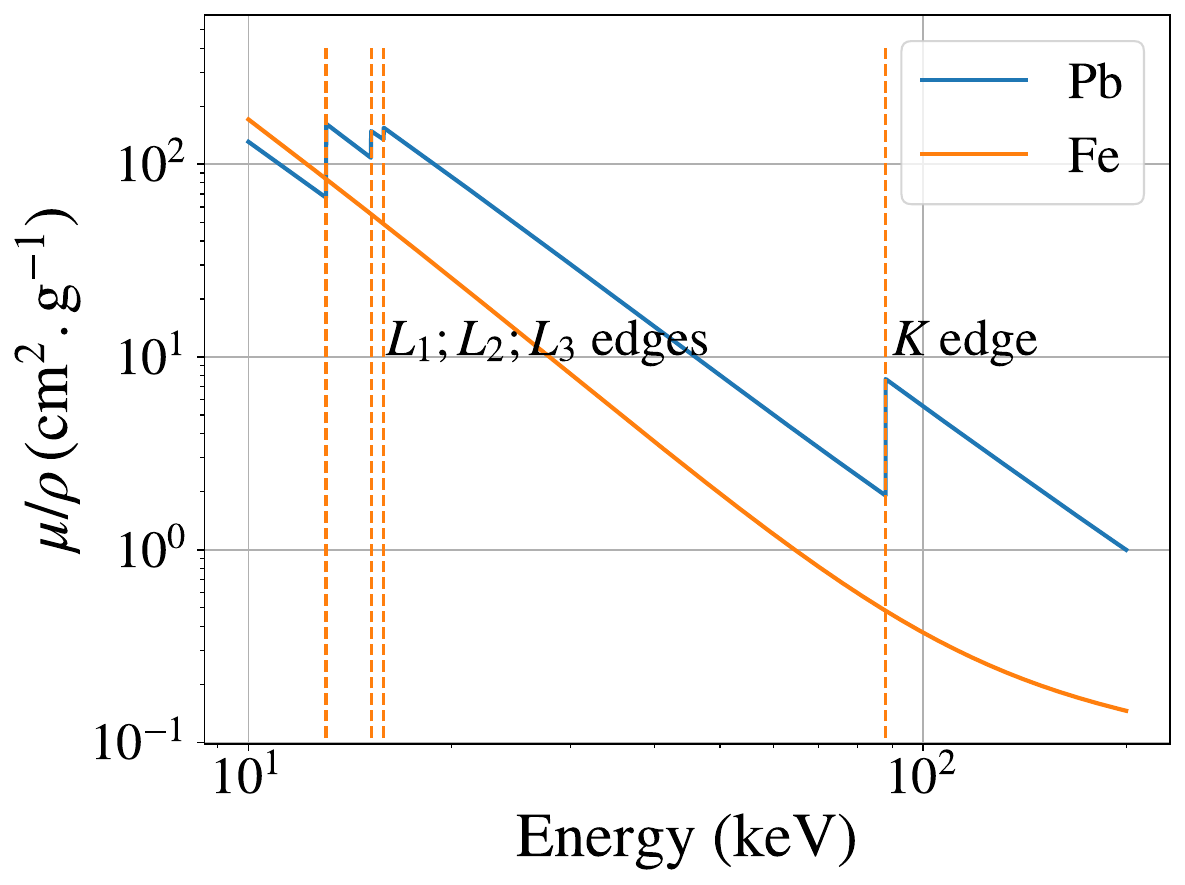}
	\end{center}	\caption{\label{fig:lead_attenuation} Attenuation coefficient of lead and iron.}
\end{figure}

Prolonged exposure to X-rays can result in significant adverse effects. In France, the annual permissible dose for a citizen is set at 1 mSv over a 12-month period. 
To comply with French regulations, the X-ray dose rate in public areas must be maintained below 80 $\mu$Sv.month$^{-1}$ or 0.5 $\mu$Sv.h$^{-1}$.
In the absence of safety measures, the dose rate around the source measures 2.0 mSv.h$^{-1}$ per hour at a distance of 1 meter. In the beam axis, this rate increases to 10.2 Sv.h$^{-1}$ at the same distance, indicating that the lethal dose can be reached within 36 minutes. 
Lead is an effective material for attenuating X-rays due to its high atomic number, which results in a significant absorption rate of X-rays, as demonstrated by its mass attenuation coefficient shown in  figure~\ref{fig:lead_attenuation}).

The K absorption edge of lead occurs at 88 keV, where its mass attenuation coefficient experiences a significant increase from 1.9 cm$^2$.g$^{-1}$ to 7.7 cm$^2$.g$^{-1}$.  
This sharp rise in the coefficient explains why lead is such an effective material for attenuating high-energy X-rays.

\begin{figure}[h]
	\begin{center}
	\includegraphics[width=0.75\textwidth]{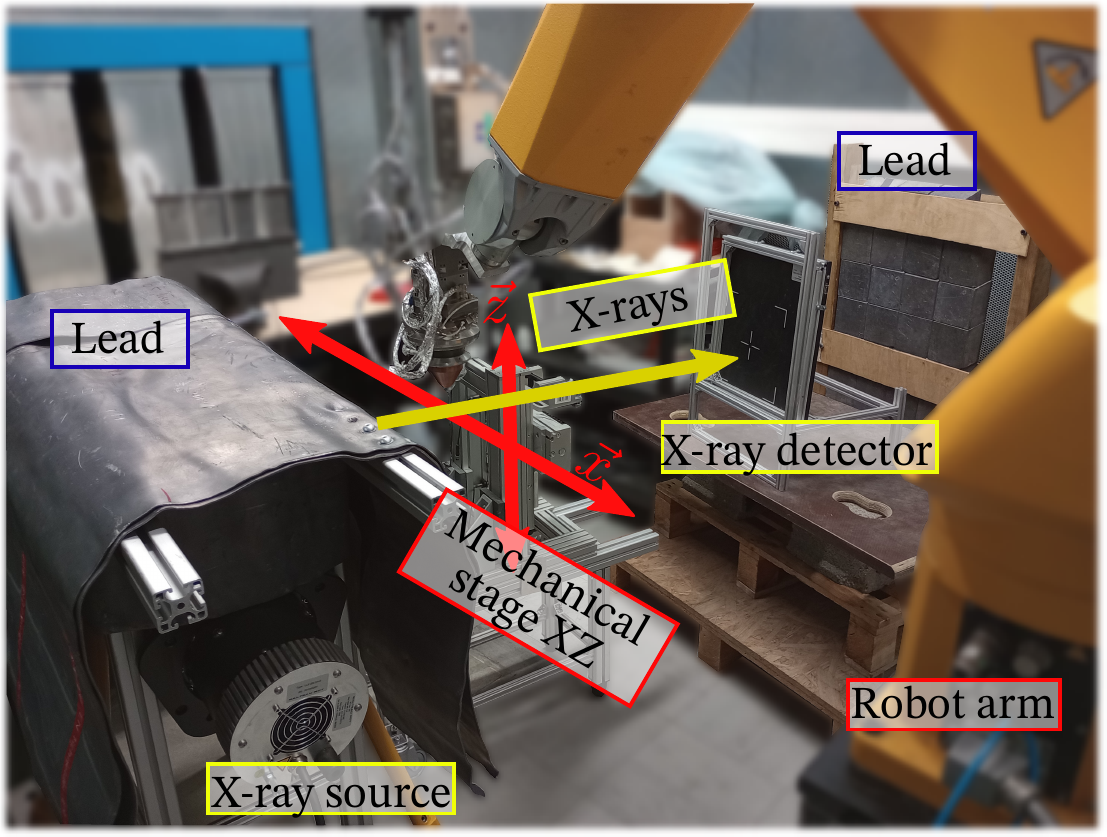}
	\end{center}	\caption{\label{fig:cabin} Experimental setup within the additive manufacturing cabin.}
\end{figure}

By covering the X-ray source with lead sheets of 4 mm thickness, the leakage dose rate is significantly reduced. Additionally, a thick wall of 50 mm lead is constructed along the beam axis to attenuate most of the primary beam (figure~\ref{fig:cabin}). 
With this shielding, the dose rate outside the additive manufacturing cabin remains below the permissible level for public areas.

\subsection{Choice of the X-ray functioning parameters}
\label{susec:choice_params}
This section explores the optimal parameters for the voltage ($U$ in kV) and intensity ($I$ in mA) of the X-ray source. To evaluate and rank different parameter sets, two indicators of image quality are employed: the signal-to-noise ratio (SNR; equation~\ref{eq:SNR}) and the contrast-to-noise ratio (CNR, equation~\ref{eq:CNR}).
The SNR is defined as follows:
\begin{equation}\label{eq:SNR}
	SNR = \frac{S_i}{\sigma_i},
\end{equation}
where $S_i$ is the mean signal in an homogeneous area $i$ of the image (typically measured in gray level), and $\sigma_i$ denotes the standard deviation of the signal within the same area (with the same units as $S_i$). A higher SNR indicates that the noise is less significant. Measuring the SNR is straightforward and can be accomplished prior to the experiment by positionning a 2 mm thick steel plate (which is made from the same material and is approximately the same thickness as the samples under study) in front of the X-ray source.

\begin{figure}[h]
	\begin{center}
	\includegraphics[width=\textwidth]{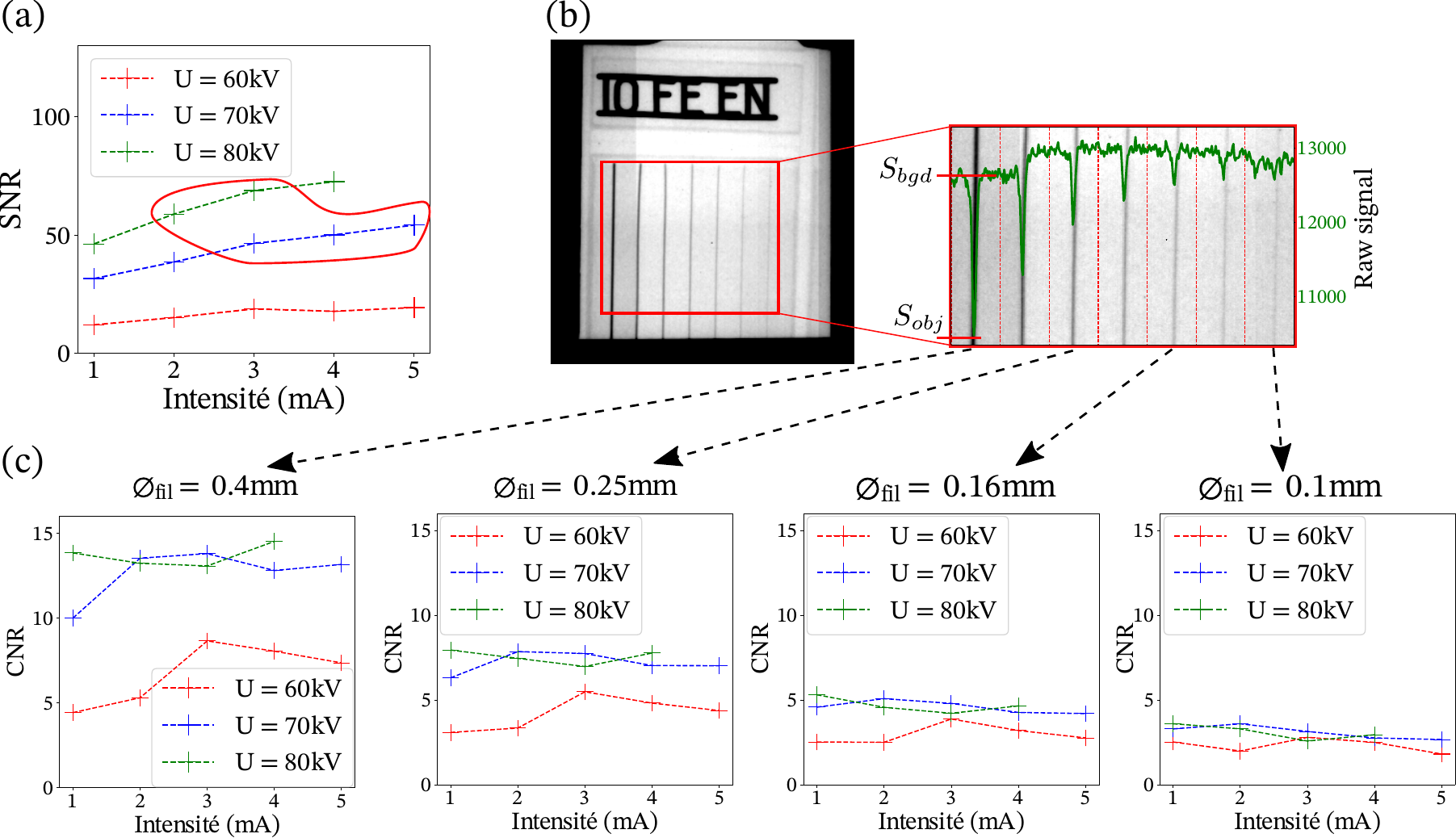}
	\end{center}	\caption{\label{fig:iqi} (a) Evolution of the SNR. (b) Radioscopy of the wire type images quality indicators and (c) the evolution of the CNR.}
\end{figure}

The CNR is defined as follows:
\begin{equation}\label{eq:CNR}
	CNR = \frac{|S_{bck} - S_{obj}|}{\sigma_{bck}},
\end{equation}
where $S_{bck}$ and $S_{obj}$ are respectively the mean signal in the background and the studied object, while $\sigma_{bck}$ indicates the associated standard deviation. In this case, the area of the object is too small to obtain a reliable standard deviation; thus the CNR is only normalized by $\sigma_{bck}$. A higher CNR value indicates that the object is more easily distinguishable from the background.
To facilitate the measurement of CNR, Images Quality Indicators (IQIs) were employed, with the wire type being used for this experiment (figure~\ref{fig:iqi}~(a)).

Seven low-alloy steel wires are aligned in decreasing diameters (from left to right: $\emptyset_{wire}$ = [0.400, 0.320, 0.250, 0.200, 0.160, 0.125, 0.100] mm). In industrial radioscopy, this IQI is positioned above the surface being studied, closest to the X-ray source and furthest from the detector. Here, the IQI is used prior to the experiment on a 2 mm stainless steel plate.
Both the CNR and SNR were evaluated as functions of the X-ray source's voltage and intensity (figure~\ref{fig:iqi}).

The five points circled in red in figure~\ref{fig:iqi}~(a) represent the highest SNR values. At 80 kV and 4 mA (the highest set of parameters), the signal the maximum value for 14-bit encoded images (approximately $15 000$ out of a maximum of $2^{14}-1=16 383$), risking overexposure in certain areas. On the contrary, at 60 kV across all intensities and at 1 mA for all voltages, the signal is too low (below $2 500$), leading to potential underexposed areas.
For each wire diameter, the CNR peaks at voltages of 70 and 80 kV and remains consistent for intensities above 2 mA (figure~\ref{fig:iqi}~(c)). For safety considerations, the lowest set of parameters is selected: 70 kV and 3 mA.
The expected CNR due to the density variation of SS316L is relatively low. According to Beer-Lambert's law, a 12\% variation in density has the same impact on the signal intensity as a 12\% variation in the thickness of the material being studied. In that case, $0.12 \times 2=0.24$, suggesting a CNR equivalent to that of the wire IQI with a diameter of 0.25 mm, which is quite good, $CNR = 7$. However, because density variation diminishes gradually around the melt pool (from 12\% at its center to 0\% at its edges), the actual CNR will be significantly lower, making density variations difficult to identify using conventional image processing techniques. One potential solution to address this problem is to analyze the contrast in terms of thickness of material penetrated by the X-rays. According to Beer-Lambert's law, the signal on the detector depends on $exp(\frac{-\mu}{\rho}\times \rho x)$. The massic attenuation coefficient remains constant for a specific material. Thus, the effects of density and thickness on signal intensity are equivalent. By assuming a constant thickness, it becomes possible to compute the density variations of the studied object. The opposite is also true, by assuming a constant density, thickness variations can be computed. For calibration purposes, since studying thickness variations in SS316L is more straightforward, the latter approach will be prioritized, and the experimental setup will be calibrated to estimate the thickness of an object based on the signal measured in the radioscopy.

%================================================================

\section{Acquisition and analysis of \textit{in-situ} radioscopy}\label{chap:acquisition}

All the experimental devices have been described in the previous section. X-ray experiments are conducted with the appropriate safety measures, and the acquired images are analyzed to investigate the shape of the melt pool.

\subsection{Image acquisition}
For this experiment, ten layers of SS316L (the choice of this material is discussed at the end of this paper \ref{sec8}), are successively deposited onto a substrate. The fabrication parameters are as follows: 
\begin{enumerate}
	\item laser power, 300 W.
	\item horizontal velocity of the mechanical stage, 3 mm.s$^{-1}$.
	\item length of a layer, 50 mm.
	\item break duration between the deposition of two successive layers, 30 s.
	\item vertical displacement of the mechanical stage between two successive layers, 0.2 mm.
	\item carrier gas flow rate (argon), 6 L.min$^{-1}$.
	\item powder flow rate, 3.4 g.min$^{-1}$.
\end{enumerate}

Compared to classical fabrication processes, the primary difference in this experiment is the break duration between two layers, which typically lasts a few seconds. In this case, a break of 30 seconds is necessary to ensure two things: to allow sufficient time for transferring the image from the detector to the computer and to avoid any afterglow effects on the detector. Indeed, the scintillator in the detector requires time to reset after an exposure to X-rays.
This extended break duration alters the thermal cycle of the part being built, with heat accumulation in the substrate being less impactful on the initial layers. In the future, this 30 seconds break will be avoided by installing a more recent x-ray detector.

\begin{figure}[h]
	\begin{center}
	\includegraphics[width=.9\textwidth]{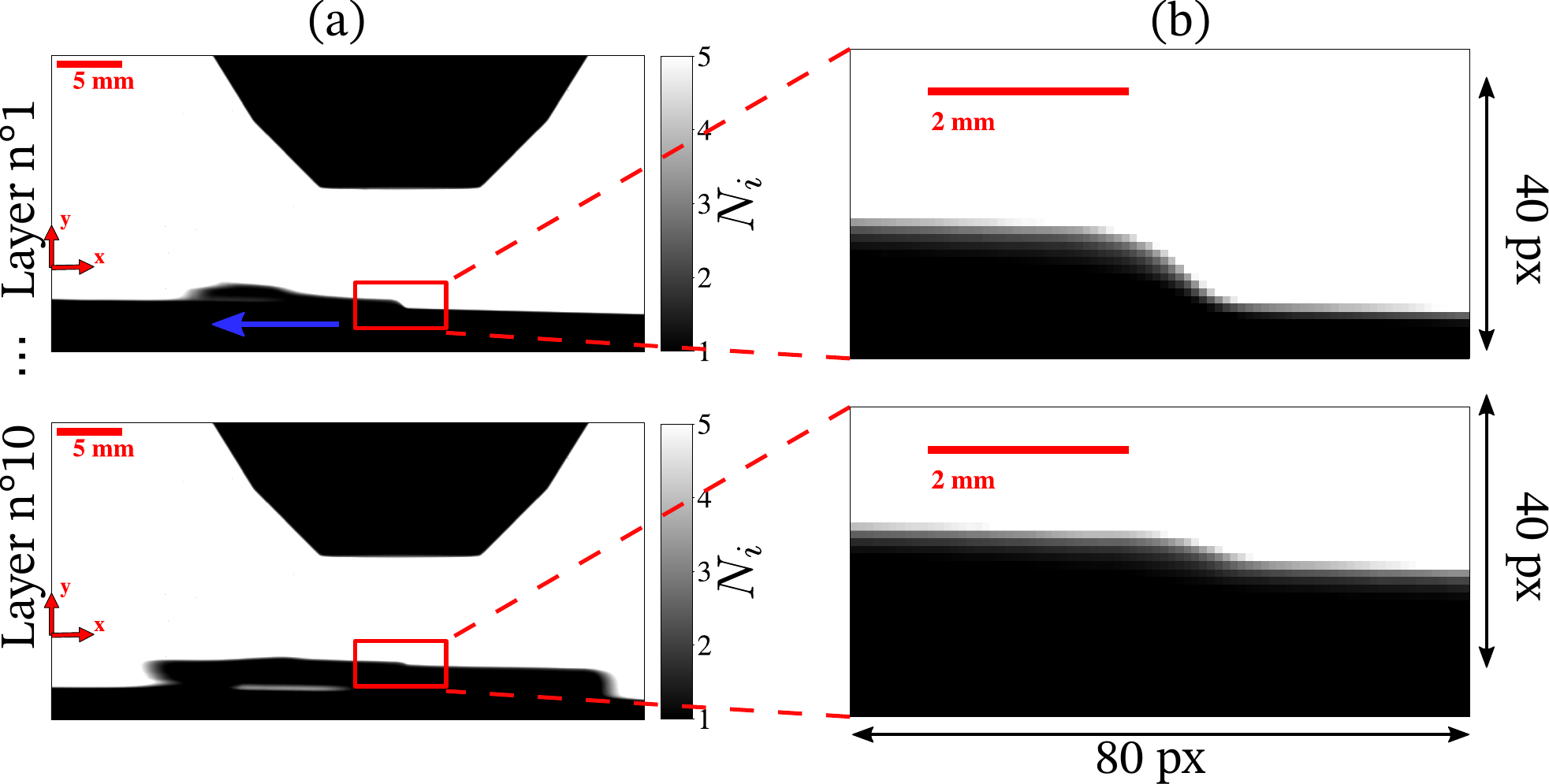}
	\end{center}	\caption{\label{fig:acqu} (a) Whole radioscopy of the laser and the layer being deposited on the first and tenth layers. (b) Zoom in under the laser.}
\end{figure}

With these parameters, it is possible to make one acquisition of a radioscopy per layer (figure~\ref{fig:acqu}). Zooming in under the laser reveals a slight bump, indicating the newly deposited layer and the presence of the melt pool.

\begin{figure*}[h]
	\begin{center}
	\includegraphics[width=\textwidth]{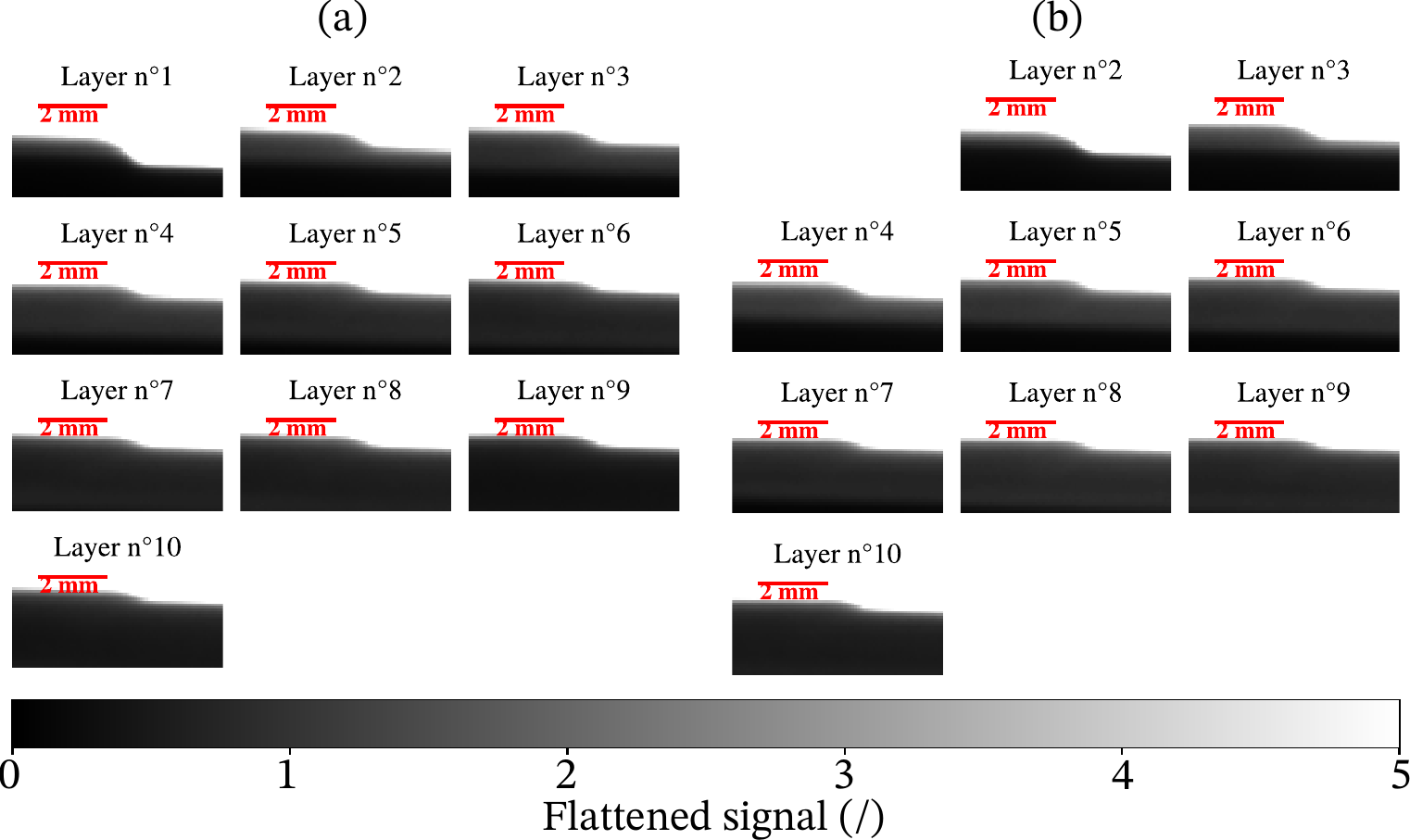}
	\end{center}	\caption{\label{fig:all_acqu} All 10 radioscopies of (a) the first experiment and (b) the second experiment.}
\end{figure*}

Two experiments were conducted under identical conditions to compare the results (figure~\ref{fig:all_acqu}~(a) and (b)). In the second experiment, the acquisition started after the first layer, which is therefore absent from the data.

In both sets of images, the height of the deposited layer decreases as the number of layers increases. Notably, the height of the bump is nearly three times higher in the first two layers (layers 1 and 2) compared to the last two layers (layers 9 and 10). This behavior can be attributed to a well-known self-regulation phenomenon in additive manufacturing processes \cite{Haley2019}.
The objective of analyzing these images is to identify the contours of the melt pool, which should appear lighter due to the variation in density. However, no apparent gray level variations are discernible to the naked eye, prompting the use of image analysis algorithms to process these images.

\subsection{Image analysis}
\begin{figure}[h]
	\begin{center}
	\includegraphics[width=.75\textwidth]{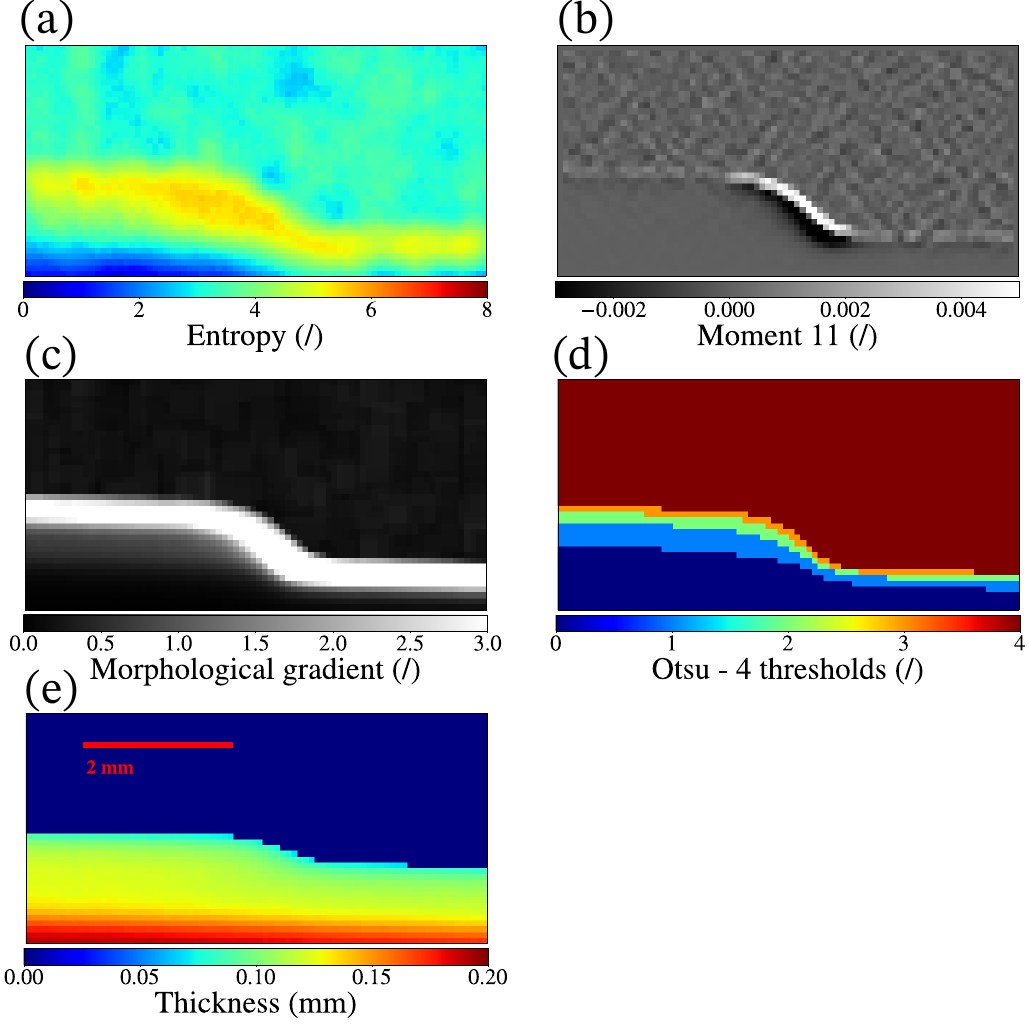}
	\end{center}	\caption{\label{fig:im_anal} (a) Entropy. (b) Moment of order 11. (c) Morphological gradient. (d) Otsu's method with 5 thresholds. (e) Thickness.}
\end{figure}

A wide variety of image analysis techniques can be employed depending on the specific case of study. In this context, the radioscopies are presented in gray scale, and the objective is to highlight low-contrast areas. Four methods have been studied:
\begin{enumerate}
	\item the entropy (figure~\ref{fig:im_anal}~(a)). This method provides information about the amount of information contained in an image. A blank image would have an entropy close to 0, while a well-defined outline will exhibit a higher entropy. It is computed via the \textit{skimage} library \cite{Walt2014}.
	\item The moment of order $pq$ ($M_{pq}$, figure~\ref{fig:im_anal}~(b)) that on an image $I$ of dimensions $n\times m$ pixels with $x$ and $y$ the pixel coordinates ($(x \leq n,y \leq m)\in \mathbb{N}$) is computed as a function of the $k$ nearest neighbors: 
			\begin{equation}\label{eq:moment}
		M_{p,q}(x,y) =  \sum_{i=x-k}^{x+k}\sum_{j=y-k}^{y+k}(i^p j^q I(i,j)).
		\end{equation}
		
	\item The morphological gradient (figure~\ref{fig:im_anal}~(c)), that is the difference between the dilation and erosion of a given image, corresponding to the \textit{dilate} and \textit{erode} functions in the \textit{OpenCV} library \cite{itseez2015opencv}).
	\item Otsu's thresholding (figure~\ref{fig:im_anal}~(d)), that is an iterative algorithm that automatically determines optimal thresholds within a given image \cite{Xu2011}. 
\end{enumerate}

Finally, a fifth method consists in computing the thickness of the studied part (figure~\ref{fig:im_anal}~(e)), by calibrating the experimental setup (x-ray source and detector) beforehand (\ref{susec:calib_expe}).
In this particular image, the computed thickness of the manufactured layers appears less than 2 mm. This discrepancy arises because the last layer takes on a semi-cylindrical shape, resulting in a consistently reduced thickness at the top of the layer. 
This calibration method assumes that the material being analyzed is homogeneous and possesses a constant density. However, this assumption does not hold true in the case of the melt pool, which is liquid and therefore less dense. Consequently, the variation in density manifests as a thickness variation in the computed results. This method is particularly valuable as it is physics-based, modifying pixel values in accordance with Beer-Lambert's law.

None of these five methods are sufficient to clearly highlight the contours of the melt pool and facilitates its identification. The entropy and the moment of order 11 both emphasize the bump, but this is primarly due to the 45-degree contour between the air and the solid part. Similarly, the morphological gradient and Otsu's thresholding detect the separation between the air and the solid material, but the contrast of the melt pool remains too low to be discernible. 
While computing the thickness of the part provides valuable insights, any potential variations in thickness within the melt pool area are obscured by the natural geometric variations of the layer, which can span several hundred micrometers. Besides, it is exacerbated by the surface tension of the deposited layers, which tend to form a spherical shape, resulting in increased thickness at their base.

Finally, the contrast between the melt pool and the surrounding material is too subtle to be identified using conventional image analysis techniques. More advanced processing techniques have been suggested, such as a template-bayesian based approach by \citet{lindenmeyer2021template} to segment the melt pool on synchrotron absorbtion images. Despite showing promising results, this method also relies on an edged segmentation that is inefficient on the current data. However, it suggests another way of solving the problem by improving the experimental setup, with a more suitable detector. For instance, a higher frame rate would allow a better preprocessing of the images, by studying the local gray level variations between two successive pictures.
This is where AI steps in. In recent years, advancements in computer vision (a field of science focusing on enabling computers to interpret and process images) have kept progressing endlessly, especially in the medical domain, which frequently resorts to X-ray imaging and faces similar challenges: identify areas that are indistinguishable to the naked eye.

%================================================================

\section{Dataset creation}\label{sec:simulation}
The goal of this section is to generate a dataset that contains radioscopies of the melt pool, associated with labels that represent the shape of the melt pool. This dataset will be used to train multiple neural networks aimed at identifying the melt pool in low-contrast X-ray images.
The simulation of the radioscopies is conducted in three stages : 
\begin{enumerate}
	\item the thermal simulation of a laser moving onto a non-planar surface,
	\item the discretization of these thermal results in terms of density (reflecting the changes in material properties due to phase transitions),
	\item the simulation of an X-ray shot directed at the density-discretized results.
\end{enumerate}

The first and last steps of this process are the most expansive in terms of time and computational resources. To streamline the simulations and optimize computational efficiency, several simplifying assumptions are made about the laser source, the x-ray source and the material properties. All these aspects are discussed in details in the appendix (\ref{secA1}). 

\subsection{Visualisation of the dataset}
In AI, a dataset refers to a collection of data used to train and test algorithms. In this study, the dataset consists of two main components: 
\begin{enumerate}
	\item an input, a simulation of a radioscopy of the melt pool (generated through the previously described methods),
	\item associated with a label, a binary image of the melt pool (derived from the thermal simulation by assuming that every voxel with a temperature exceeding 1673 K corresponds to the melt pool).
\end{enumerate}

\begin{figure}[h]
	\begin{center}
	\includegraphics[width=8cm]{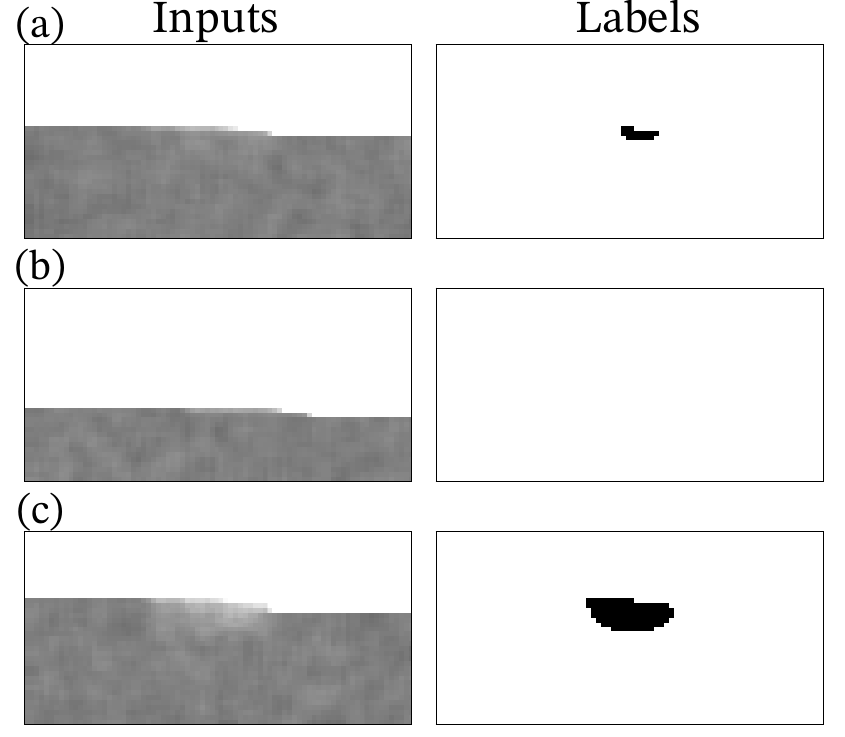}
	\end{center}	\caption{\label{fig:dataset} Simulation of the radioscopy, the thermal simulation and the associated binary label : (a) with a small melt pool, (b) without melt pool ($P_{laser} = 100$ W and $v_{laser} = 9$ mm.s$^{-1}$) and (c) with a big melt pool and an offset.}
\end{figure}

A total of 162 thermal and X-ray simulations are performed, with laser power varying between 100 and 950 W and velocities ranging from 1 to 9 mm.s$^{-1}$. At the lowest powers and highest velocities, the heat input is insufficient to create a melt pool (figure~\ref{fig:dataset}~(b)). However, these data are valuable for training the model to recognize scenarios where no melt pool is present in the X-ray image. The proportion of images without melt pool accounts for 15\% of the dataset.
To enhance the dataset further and avoid the melt pool always being centered in each image, a random offset is added to the melt pool position (figure~\ref{fig:dataset}~(a) and (c)).

At this point, the dataset is completely created, made of 4050 images, and ready to be used to train and test the convolutional auto-encoders.

\subsection{Convolutional auto-encoder}
\begin{figure}[h]
	\begin{center}
	\includegraphics[width=0.5\linewidth]{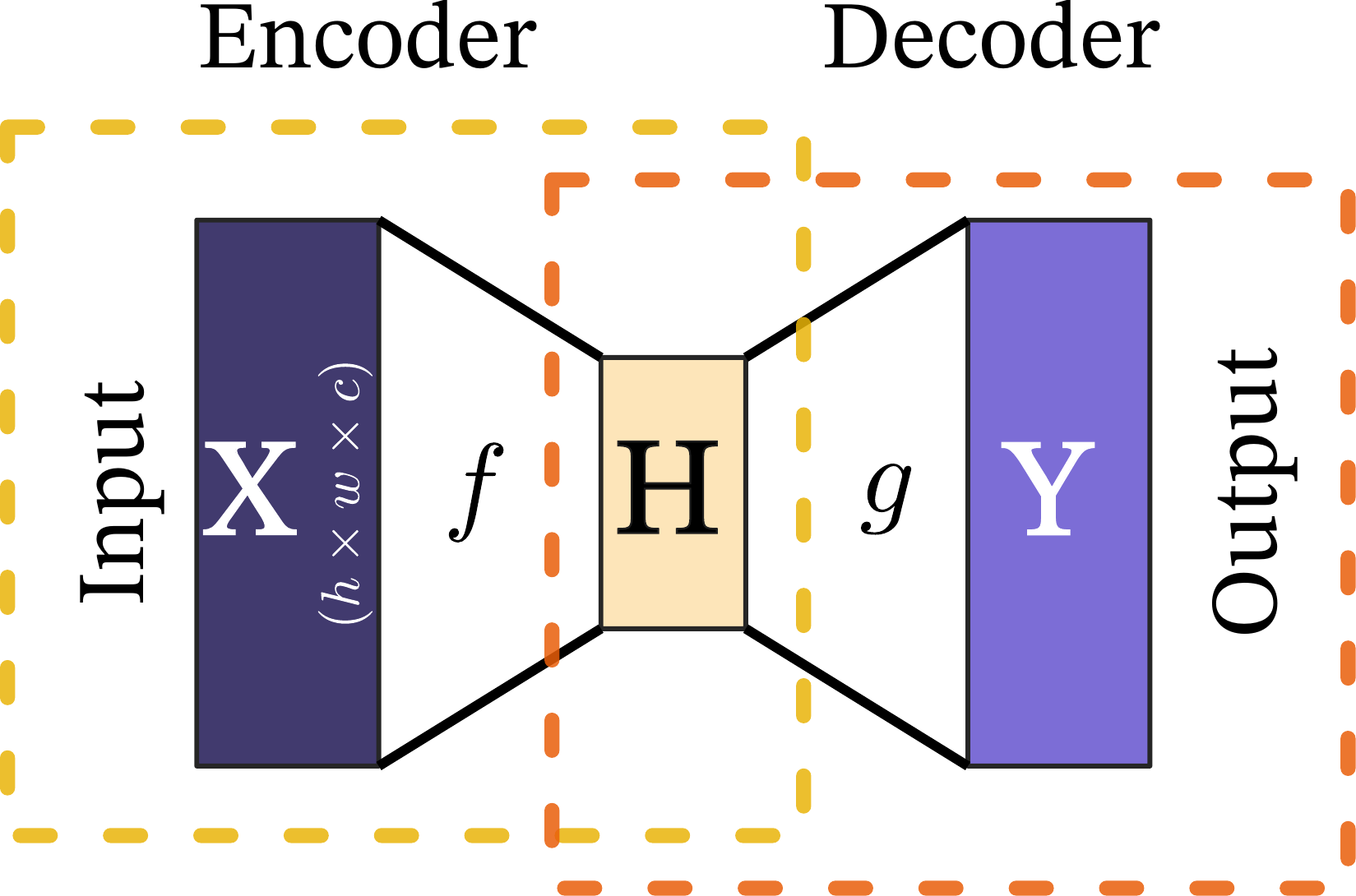}
	\end{center}	\caption{\label{fig:AEC} Simplified representation of an auto-encoder.}
\end{figure}

A convolutional auto-encoder is a type of neural network that consists of two back-to-back CNN: an encoder and a decoder (figure~\ref{fig:AEC}). The input is an image $X$ with dimensions ($h\times w \times c$), where $h$ represents the height, $w$ the width, and $c$ the number of color channels (for a standard \textit{Red-Green-Blue} image, $c=3$). The image is processed by the encoder $f$, which converts the input into a new representation $H$ (typically a compressed version capturing the main features of the input based on the training). This representation is then passed through the decoder $g$, resulting in a new output image $Y$ with the desired dimensions (\citet{Zhang2019} provide a more detailed review about convolutional neural networks).

\subsubsection{Layout}
\begin{figure*}[h]
	\begin{center}
	\includegraphics[width=\linewidth]{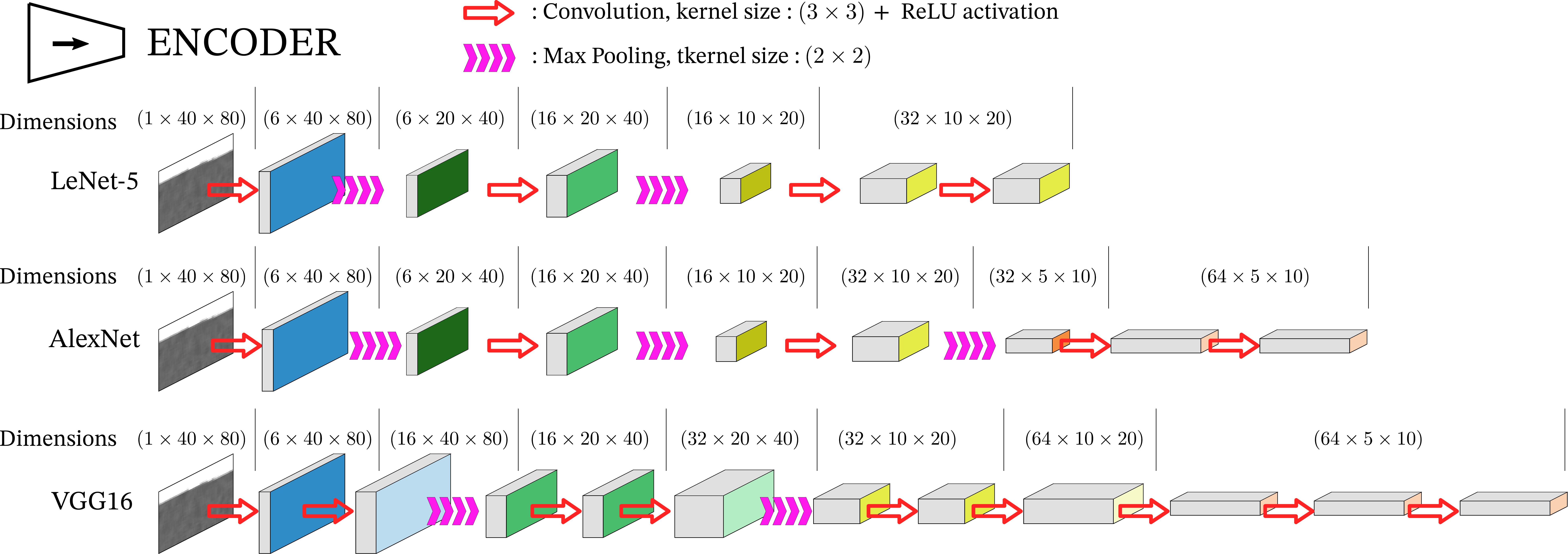}
	\end{center}	\caption{\label{fig:encoder} Architecture of the 3 auto-encoders inspired from LeNET5, AlexNET and VGG16.}
\end{figure*}

In this study, three architectures of auto-encoders have been studied, inspired by LeNET5, AlexNET, and VGG16 (figure~\ref{fig:encoder}).

The convolution layers, indicated by the red arrows, all have a stride and padding of 1 with a kernel size of $(3,3)$ (see \citet{Dumoulin2016} for a more detailed discussion on the impact of these parameters). After each of these layers, the activation function applied is the Rectified Linear Unit (ReLU) \cite{Sharma2020}. The max-pooling layers \cite{Scherer2010}, represented by the pink arrows, serve to reduce the size of the data.
In contrast, the decoders mirror the architecture of the encoders but in reverse: transposed convolutions replace the standard convolutions, and upsampling replaces the max-pooling layers.

\subsubsection{Training process}
The training process involves adjusting the parameters of the layers to produce an output that aligns with the labels corresponding to given inputs. First, the dataset is divided into a training set ($70\% =2835$ inputs) and a test set ($30\% = 1215$ inputs), with the test set being completely excluded from the training process.
The training data is further divided into batches of 50 and 200 images in order to reduce the risk of overfitting and speeding up the optimization of the auto-encoder. It is important to make the difference between an epoch and an iteration; for example, with $1 000$ images in the training set and a batch size of $200$, completing one epoch requires 5 iterations.
To quantify the auto-encoder's error, the output ($x_i$) is compared with the associated  label ($y_i$) using the \textit{Mean Squared Error} (MSE, equation~\ref{eq:MSE}): 
\begin{equation}\label{eq:MSE}
	\begin{aligned}
	&L_{MSE} = \{mse_1,...,mse_N\}^T, mse_n = (x_n - y_n)^2, \\
	&MSE(x,y) = mean(L),
	\end{aligned}
\end{equation}

with : 
\begin{itemize}
	\item $y_i$ the label of the input,
	\item $x_i$ the prediction of the input by the autoencoder,
	\item $N$ the number of data in the batch.
\end{itemize}
After processing each batch, the auto-encoder's coefficients are updated, and the results are recorded over the course of 600 epochs. 

\subsubsection{Results and loss}
\begin{figure}[h]
	\begin{center}
	\includegraphics[width=8cm]{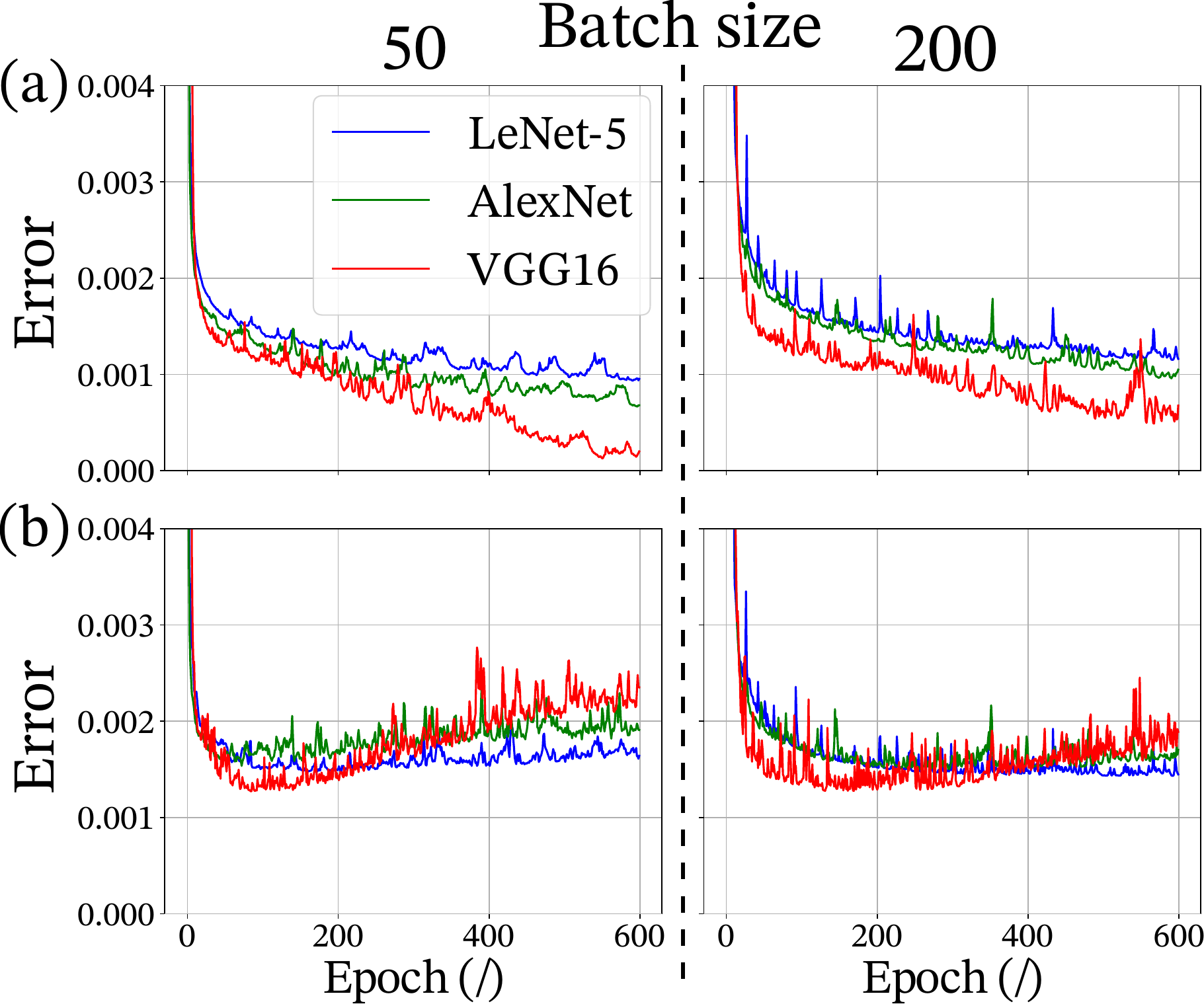}
	\end{center}	\caption{\label{fig:loss} (a) Training loss. (b) Generalization loss.}
\end{figure}

At the end of each epoch, the mean loss on the training set (referred to as the training loss, figure~\ref{fig:loss}~(a)) and the mean loss on the test set (referred to as the generalization loss, figure~\ref{fig:loss}~(b)) are recorded. An effective neural network strikes a balance between these two error metrics.

For all architectures, the training loss consistently decreases as the number of epochs increases. On the opposite, the generalization loss also rapidly decreases during the first hundred epochs, after which it reaches a minimum and begins to increase gradually. This increase indicates signs of overfitting. The choice of the error (whether MSE or Huber) does not significantly affect the performance evaluation of the auto-encoder in this study. Additionally, the batch size influences both the training and generalization loss; larger batch sizes tend to raise the training error but result in a slightly lower generalization error, which is more desirable to reduce risks of overfitting.

Overall, these trends suggest that the problem being addressed is not particularly challenging, as all three architectures demonstrate comparable performance. The size of the saved file after training each autoencoder serves as a useful indicator of their complexity (table~\ref{tab:savefile}).

\begin{table}[h]
\caption{Size of the saved files for every auto-encoder.}\label{tab:savefile}%
\begin{tabular}{@{}ll@{}}
\toprule
Auto-encoder & Size of the saved file (kb) \\
\midrule
LeNET5 & 58\\
AlexNET & 504\\
VGG16 & 1036\\
\botrule
\end{tabular}
%\footnotetext{Source: This is an example of table footnote. This is an example of table footnote.}
%\footnotetext[1]{Example for a first table footnote. This is an example of table footnote.}
%\footnotetext[2]{Example for a second table footnote. This is an example of table footnote.}
\end{table}

%    \begin{table}[pos = H]
%      \centering
%      \begin{tabular}[x]{|m{7em}|m{7em}|}
%      \hline
%      \textbf{Auto-encoder} & Size of the saved file (Ko) \\
%      \hline
%       LeNET5 & 58\\
%      \hline
%       AlexNET & 504\\
%      \hline
%       VGG16 & 1036\\
%      \hline
%      \end{tabular}
%      \caption{Size of the saved files for every auto-encoder.}
%      \label{tab:savefile}
%    \end{table}
    
\begin{figure}[h]
	\begin{center}
	\includegraphics[width=.75\textwidth]{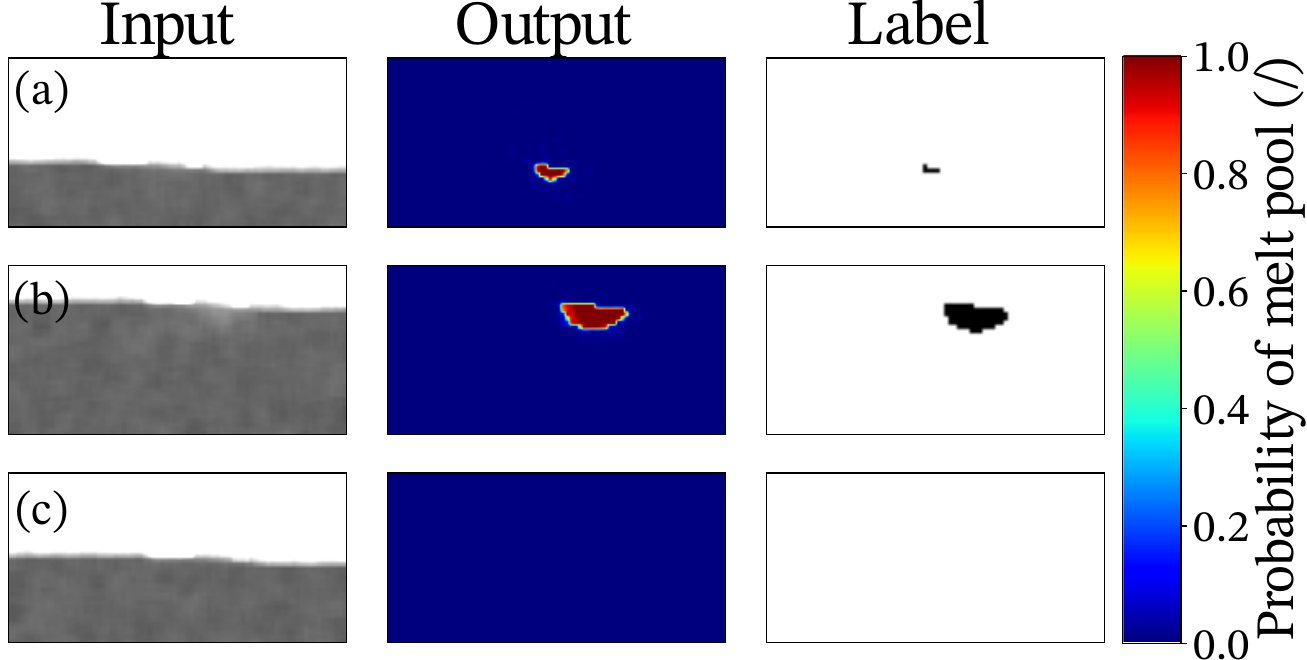}
	\end{center}	\caption{\label{fig:results} Identification of the melt pool onto the simulation of the test set with the VGG16 inspired auto-encoder, with batches of 200 and trained during 400 epoch : (a) little melt pool, (b) big melt pool with an offset and (c) no melt pool.}
\end{figure}

VGG16 is the most complex architecture, with nearly twice as many parameters as AlexNET. To evaluate the performance of these models, they are applied to inputs from the validation dataset, which were not included in the training process. The results on the simulation dataset are quite promising (figure~\ref{fig:results}). 

The color legend on the output can be interpreted as the probability that a particular pixel belongs to the melt pool. In cases where there is no melt pool (figure~\ref{fig:results}~(a)), a bigger one (figure~\ref{fig:results}~(b)) or a medium-sized one (figure~\ref{fig:results}~(c)), the auto-encoder accurately draws its shape.
The next step is to evaluate how effectively these CNNs, trained on simulated data, perform on experimental radioscopies.

\section{Generalization to experimental datas and discussions}\label{sec8}

\begin{figure}[h]
	\begin{center}
	\includegraphics[width=.75\textwidth]{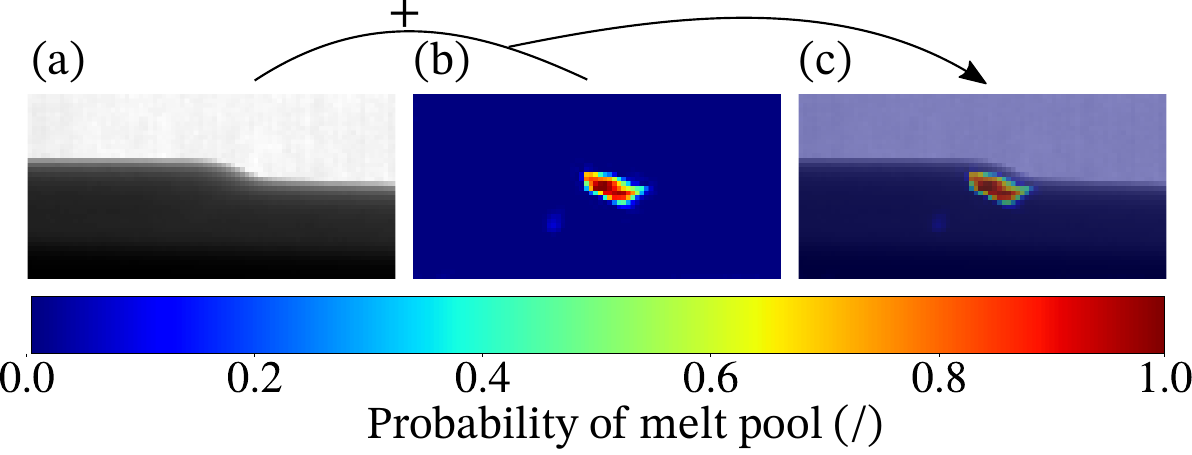}
	\end{center}	\caption{\label{fig:res_IA_0} (a) Flattened radioscopy (input of the auto-encoder). (b) Output of the auto-encoder. (c) Superimposition of (b) onto (a) for a better visualization.}
\end{figure}

The primary objective of this study is to identify the melt pool in very low-contrast in-situ radioscopies. The previous results demonstrate that the three architectures trained on simulated data effectively detect the melt pool, regardless of its size and position within the simulated images. This section aims to assess how well these models generalize to experimental images.
Currently, transfer learning is a prominent area of research that involves training large neural networks on supercomputers to perform \textit{source tasks}. These same networks are then partially retrained on more accessible hardware for \textit{target task}. This approach significantly speeds up the training process compared to starting from scratch while achieving similar performance levels.
In this study, the source task involves identifying the melt pool in simulated radioscopies, while the target task focuses on identifying the melt pool in experimental images. The problem is that retraining the auto-encoders on experimental data is impractical, as it would require labeled data to identify the melt pool before training (which is not possible, figure~\ref{fig:im_anal}).
Thus, the auto-encoders trained on simulated data were applied directly to the experimental images without any modifications. All flattened radioscopies obtained during the experimental tests serve as inputs (figure~\ref{fig:all_acqu}). 
To visualize the results, the outputs of the auto-encoders are overlaid on the original radioscopies (figure~\ref{fig:res_IA_0}).

The three architectures (LeNET5, AlexNET, and VGG16) are evaluated using the same dataset. Each model was trained on this dataset with batches of 200 images over the course of 400 epochs.

\begin{figure}[h]
	\begin{center}
	\includegraphics[width=\textwidth]{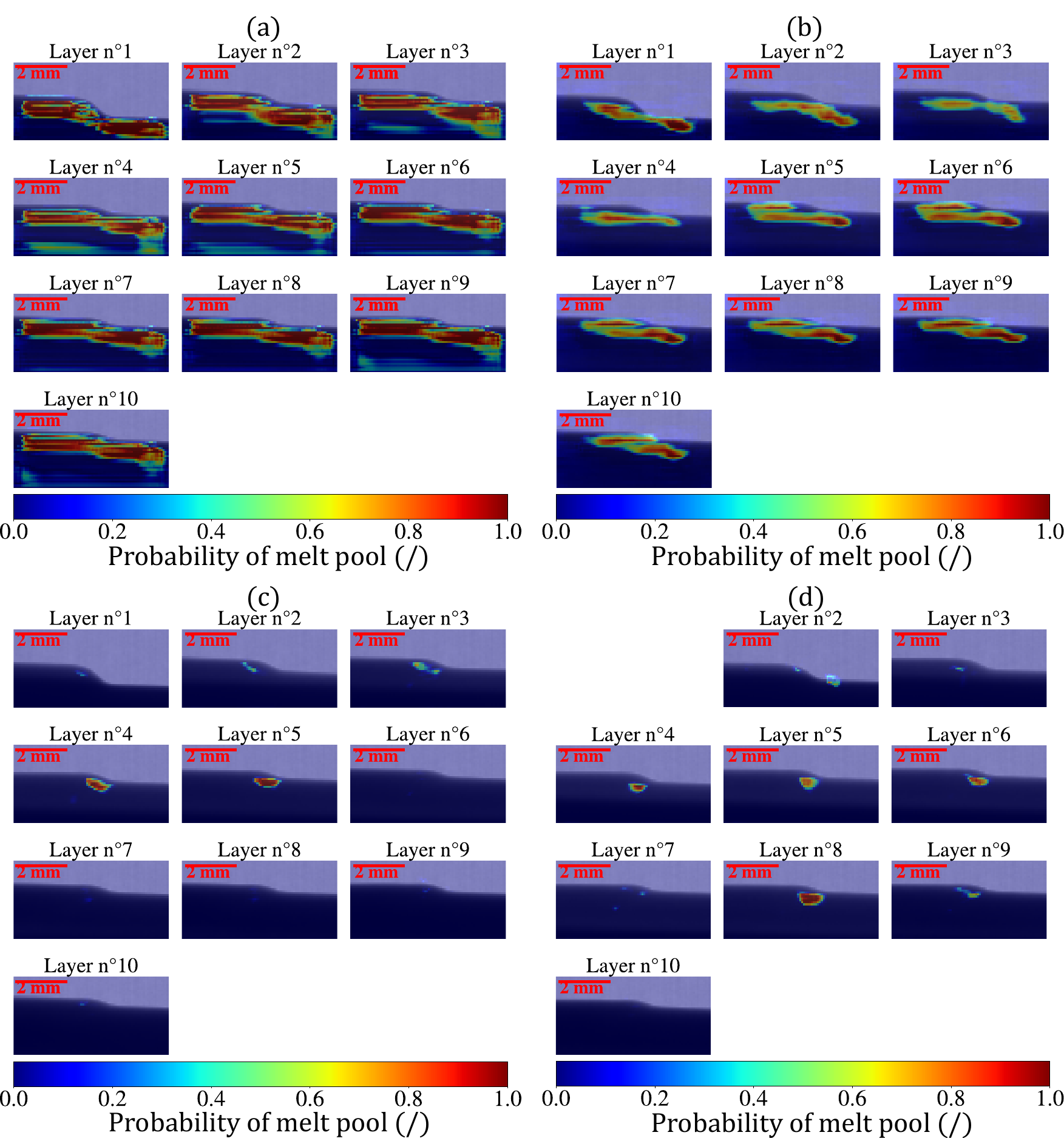}
	\end{center}	\caption{\label{fig:res_IA_1} Results of the three architectures on the experimental radioscopies (batch size 200, 400 epochs). First experiment : (a) LeNET5, (b) AlexNET and (c) VGG16. Second experiment (d) VGG16.}
\end{figure}

The initial two architectures, LeNET5 and AlexNET (figure~\ref{fig:res_IA_1}~(a) and (b)), do not succeed in identifying the contours of the melt pool. The colored areas appear scattered around the expected location of the melt pool, which is situated beneath the small bump. Extending the training duration for these architectures is unlikely to improve the generalization error. 
On the contrary, the VGG16-inspired auto-encoder (figure~\ref{fig:res_IA_1}~(c) and (d)) demonstrates more promising results. In the first experiment (figure~\ref{fig:res_IA_1}~(c)), it successfully identifies the melt pool and locates it correctly within layers 3, 4, and 5. However, in that case, there seems to be little distinction among all the images. 
In the second experiment (figure~\ref{fig:res_IA_1}~(d)), similar results are observed, with the melt pool detected directly beneath the bump in layers 4, 5, 6, and 8.
Currently, there is no straightforward method to verify whether the identified melt pools have accurate shapes. One potential approach is to utilize a thermal camera for comparison, which was developed in previous work \cite{Jegou2023}. This device could be mounted on the additive manufacturing setup to capture overhead views of the melt pool, allowing for a comparison between the dimensions of the melt pool on the surface (via thermal images) and those within the substrate (using radioscopies). A second option would be to compare the observed shapes with those calculated from a comprehensive simulation of the process, which accounts for factors such as powder supply and the movement of the melt pool; this is currently under investigation \cite{Jia2024}.
A final option could be established by pairing these results with synchrotron-based imaging, for instance on ID19 at the European Synchrotron Radiation Facility, where additive manufacturing devices are used during in-situ experiments \citep{bugelnig2024situ}. 
Also, in order to improve the contrast on the x-ray images, it would be interesting to study lighter materials, such as aluminium. On the current additive manufacturing setup, aluminium is not studied since aluminium dust is combustible and an explosion hazard. Safer techniques, such as wire arc additive manufacturing, would be an interesting option.

Finally, in light of these results, two assumptions can be made:
\begin{enumerate}
	\item the auto-encoder is not performing well enough, as the identification success rate for the melt pool is merely 37\%, despite the expectation that it should be identified in all images (this statistic is based on a limited number of samples).
	\item The shape of the melt pool may change during fabrication (due to heat accumulation in the substrate), and the auto-encoder was not trained to recognize shapes other than semi-spherical ones. 
\end{enumerate}	 

It is likely that a combination of both hypotheses holds true. The dataset being exclusively derived from simulations introduces bias during training. While the simulations are physics-based (with contrast differences stemming from coherent density variations), the shape of the melt pool remains constant due to the simplicity of the thermal simulation. It would be beneficial to enhance the dataset with a broader range of melt pool shapes. Alternatively, developing a model capable of identifying low-contrast differences in images without imposing shape constraints could be pursued. Such an auto-encoder would require substantial resources for training, including a large dataset and a more complex architecture, potentially inspired by U-NET \cite{ronneberger2015}).

To conclude, these preliminary results are encouraging, as auto-encoders trained solely on simulations can identify the melt pool in experimental images without any preprocessing.

%================================================================
%  Conclusion
%================================================================

\section{Conclusion}\label{chap:conclusion}

This work has demonstrated the feasibility of in-situ X-ray NDT on industrial additive manufacturing setups. The system consists of a detector, a movable X-ray source, and lead sheets to ensure safety concerning X-ray exposure. The source's operational parameters were optimized to maximize the SNR and the CNR. However, the radioscopies obtained with this setup exhibit low contrast, making it difficult to distinguish the density differences between the melt pool and the solid areas using conventional image analysis techniques applied to grayscale images. Nonetheless, these pioneering experiments yield promising results for the field of additive manufacturing. 
While previous studies have successfully identified porosity defects during fabrication, the real-time analysis of the melt pool has posed significant challenges. In this study, convolutional auto-encoders were trained exclusively on a dataset derived from simulation results. The thermal simulations provided temperature distributions within the melt pool, allowing for the computation of its density, which was then utilized for simulating X-ray interactions. These physics-based simulations aimed to ensure that the contrast between the liquid and solid regions would closely align with experimental results. Three distinct architectures were trained on this dataset and subsequently applied to experimental images without any modifications. The VGG16-inspired architecture yielded encouraging results, successfully identifying the melt pool in 37\% of the experimental images.

In conclusion, this work can be improved both experimentally and numerically. First, the detector employed in this study is 18 years old, and its acquisition rate is significantly lower than that of more modern detectors. Upgrading to a higher acquisition rate would facilitate the collection of multiple images during the deposition of a single layer, thereby generating more experimental data and providing deeper insights into melt pool behavior over shorter time frames. Secondly, the thermal simulation relied on substantial assumptions that greatly simplified the shape of the melt pool. Enhancing the generalization of the dataset would likely lead to improved generalization of the auto-encoders and higher success rates in identifying the melt pool.

\begin{appendices}

\section{Stainless steel density as a function of temperature}
\label{sususec:density}

\begin{figure}[h]
	\begin{center}
	\includegraphics[width=0.75\textwidth]{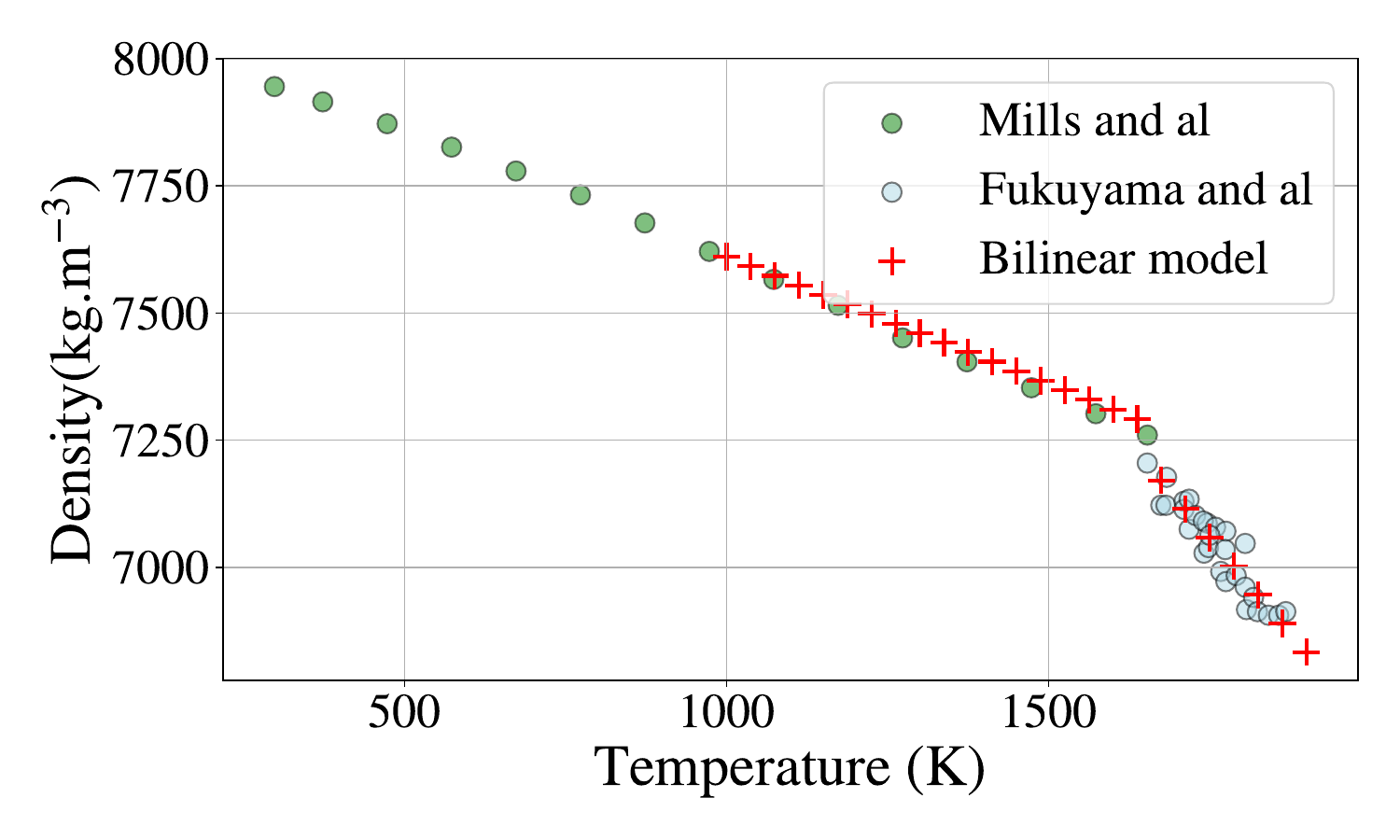}
	\end{center}
	\caption{\label{fig:density} Evolution of the density as a function of the temperature for solid and liquid SS316L.}
\end{figure}

The experiment described in this paper focus on the additive manufacturing of 316L stainless steel (SS316L). During this process, a laser is focused on a 316L plate and locally melts it. At the same time, 316L powder is delivered into this molten area. As the laser moves forward, the previously molten area solidifies, and this cycle continues. It is commonly assumed that the thickness of the produced part remains constant with LMDp. 
This study aims at observing the depth of the liquid area, specifically examining the contrast created by the density difference between solid and liquid SS316L.
The evolution of the density as a function of temperature has been extensively investigated, first in a solid state \cite{Mills2004} and then in a liquid state \cite{Fukuyama2019}. Thanks to these studies, the density of SS316L can be expressed as a function of temperature (equation~\ref{eq:densite}, figure~\ref{fig:density}).

\begin{equation}	\label{eq:densite}
	\begin{aligned}
	\rho_{316L} &= 7951.6 - 0.5T, \mbox{ if } T<1673 K \\
	&= 9683.5 - 1.5T, \mbox{ else}.
	\end{aligned}
\end{equation}

Research on this specific additive manufacturing process has indicated that the maximum temperature can approach 2000 K \cite{Jegou2023}, resulting in a density variation of approximately 12\% between the solid and liquid regions. 
Thus, this study aims at identifying the melt pool by analyzing the variation in X-ray beam intensity, which is attributed to the decrease in density of SS316L in the molten area.

%================================================================
%================================================================

\section{Calibration for thickness estimation}
\label{susec:calib_expe}
Beer Lambert's attenuation law, as stated in equation~\ref{eq:beer_lambert}, is applicable only to monochromatic energies. As a matter of fact, the massic attenuation coefficient varies with the energy of the photons; lower-energy photons are more easily attenuated by a given thickness of material. This phenomenon is referred to as \textit{beam hardening}.

\begin{figure}[h]
	\begin{center}
	\includegraphics[width=0.9\textwidth]{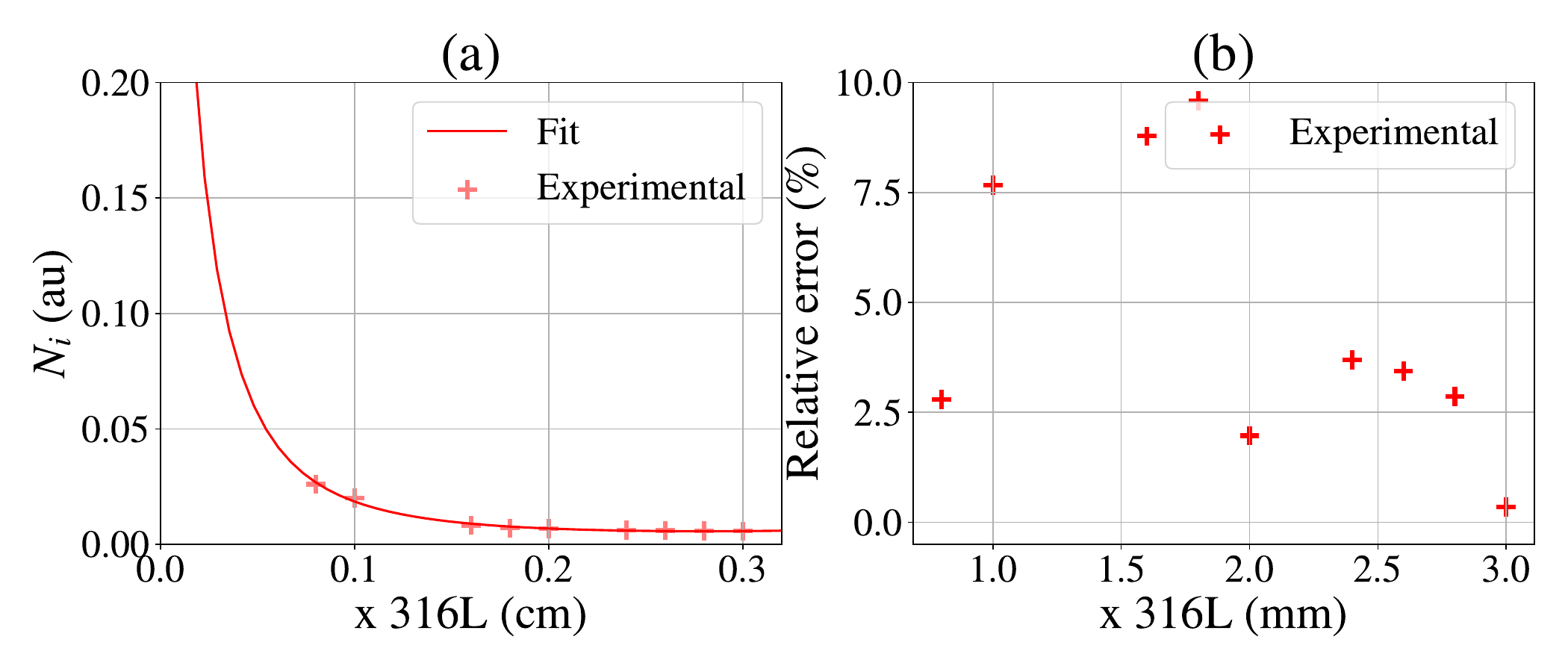}
	\end{center}	\caption{\label{fig:calib_exp} Evolution of (a) the signal and (b) the relative error on the computed thickness, as a function of the thickness of SS316L.}
\end{figure}

To calibrate the polychromatic beam for a specific material, a series of plates with known thicknesses are placed in front of the X-ray source, and the resulting signal is measured on the detector (figure~\ref{fig:calib_exp}~(a)).
This calibration is specific to a given detector in a particular configuration: 
\begin{enumerate}
	\item distance between the X-ray source and the object, and the object and the detector (respectively 680 mm and 520 mm in this experiment), 
	\item voltage and intensity of the X-ray source (respectively 70 and 3 mA in this experiment).
\end{enumerate}

For each plate, the signal received on the detector ($N$) is measured and normalized by the signal $N_0$ when a 2 mm plate is placed between the source and the detector. Although this measurement is typically performed without any plates between the source and the detector, doing so often leads to detector saturation. The 2 mm thickness corresponds to the expected thickness of the manufactured parts. This normalized signal is referred to as the flattened signal ($N_i$), which is then directly fitted using an equation inspired by \citet{Baur2023} (equation~\ref{eq:calib_exp}). 

\begin{equation}\label{eq:calib_exp}
	N_i = exp\left(ax-bx^{1-c}\right),
\end{equation}
with $a$, $b$ and $c$ the three parameters to optimize. 

\begin{table}[h]
\caption{Calibration of the setup for SS316L.}\label{tab:calib_res}%
\begin{tabular}{@{}llll@{}}
\toprule
Material & a  & b & c \\
\midrule
SS316L & 61.47 & 59.60 & 0.23 \\
row 2    & data 4   & data 5\footnotemark[1]  & data 6  \\
row 3    & data 7   & data 8  & data 9\footnotemark[2]  \\
\botrule
\end{tabular}
%\footnotetext{Source: This is an example of table footnote. This is an example of table footnote.}
%\footnotetext[1]{Example for a first table footnote. This is an example of table footnote.}
%\footnotetext[2]{Example for a second table footnote. This is an example of table footnote.}
\end{table}

%    \begin{table}[pos = H]
%      \centering
%      \begin{tabular}[x]{|m{5em}||m{3em}|m{3em}|m{3em}|}
%      \hline
%      \textbf{Material} & \textbf{a} & \textbf{b} & \textbf{c}\\
%      \hline
%       SS316L & 61.47 & 59.60 & 0.23 \\
%      \hline
%      \end{tabular}
%      \caption{Calibration of the setup for SS316L.}
%      \label{tab:calib_res}
%    \end{table}

The relative error between the measured thickness and the computed one using the calibration results (table~\ref{tab:calib_res}) is less than 10\% (figure~\ref{fig:calib_exp}~(b)). For thickness exceeding 3 mm, the measured signal is too low, as nearly all X-rays are absorbed by the material. Given that the studied objects have an approximate thickness of 2 mm, the calibration is highly effective and provides a reliable method for calculating the thickness of a homogeneous 316L object based on its signal on the detector. This is accomplished by solving equation~\ref{eq:non_linear} using a numerical method, such as the Newton-Raphson algorithm.

\begin{equation}\label{eq:non_linear}
	ax-bx^{1-c}-ln(N_i) = 0.
\end{equation}

This calibration is particularly valuable in the context of additive manufacturing, as it offers an alternative method for analyzing the acquired images. Moreover, the thickness calibration of the source can also be used to compute the density for a specified thickness, enhancing the ability to characterize the material properties of the produced parts.

\section{Dataset creation}\label{secA1}
%================================================================
\subsection{Assumptions}
The overall time of simulation (thermal and X-ray) is reduced with the following assumptions : 
\begin{enumerate}
	\item thermal properties are constant and independent from the temperature,
	\item the thermal convection loss is neglected,
	\item the laser source is Gaussian,
	\item the geometry of the melt pool is approximated by polynomial functions,
	\item the X-ray scattering is neglected, 
	\item the X-ray source is isotropic.
\end{enumerate}

\begin{figure}[h]
	\begin{center}
	\includegraphics[width=.9\textwidth]{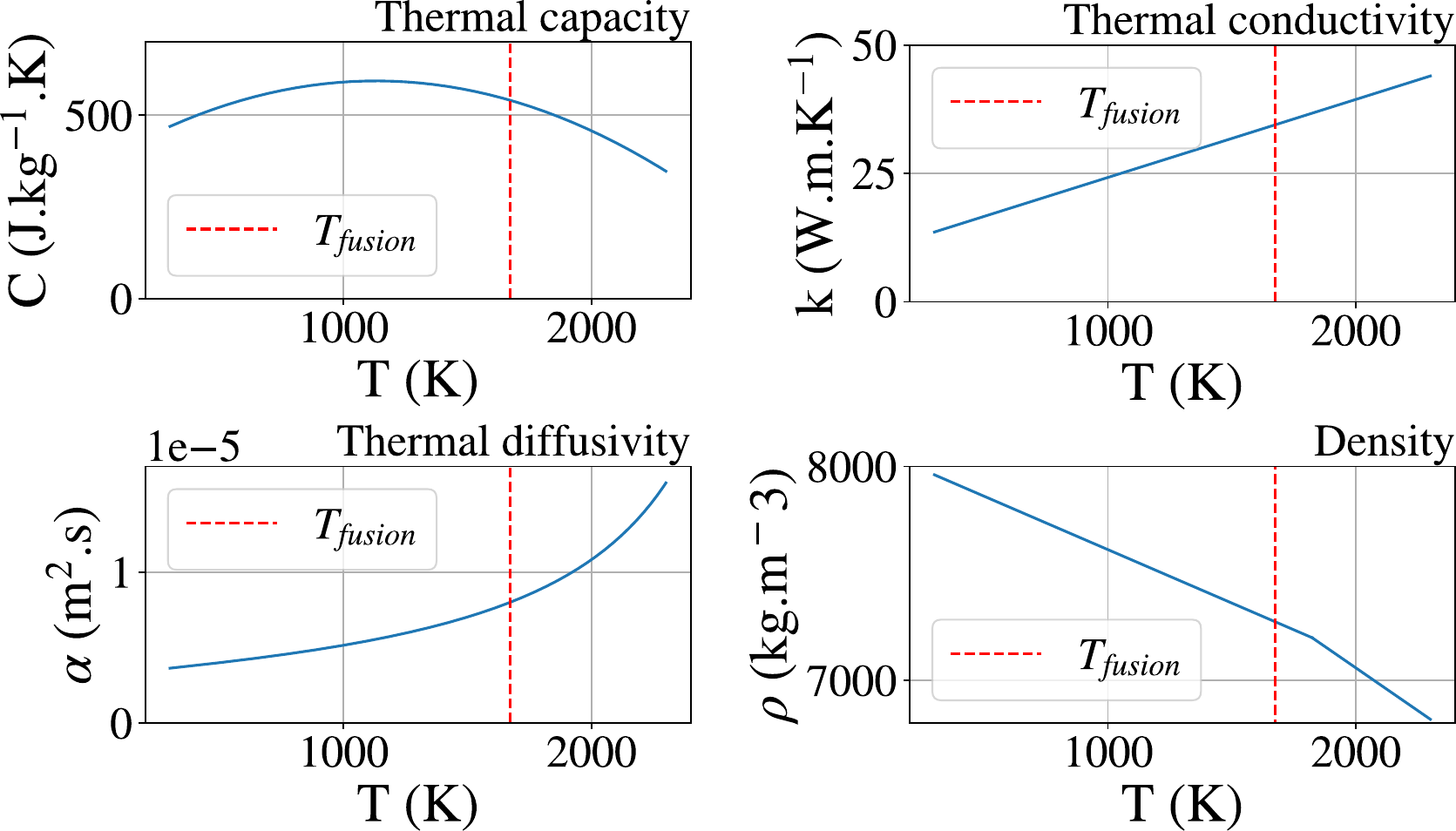}
	\end{center}	\caption{\label{fig:params} Evolution of material parameters as a function of the temperature.}
\end{figure}

The first assumption is the most constraining since the temperature equation is highly nonlinear, and every property of the material is temperature-dependent (figure~\ref{fig:params}). However, to obtain an analytic solution to the thermal problem (significantly faster than using finite element methods) it is essential to impose constant values for the thermal properties, specifically: the thermal capacity ($C = 500$ J.kg$^{-1}$.K), the thermal conductivity ($k = 20$ W.m.K$^{-1}$), the density ($\rho = 7800$ kg.m$^{-3}$) and the thermal diffusivity ($\alpha = \frac{k}{\rho C}$ in m$^2$.s).

Assumptions (2) and (3) are common in thermal simulation of additive manufacturing processes (although nowadays Goldak laser sources are also common in this domain), and the fourth simplifies the resolution of the analytic problem. 
The last two assumptions concerning the X-ray simulation are both aimed at reducing computational time and are supported by experimental evidence. When the distance between the object and the detector is large enough, the scattering of X-rays is negligible (air gap technique), and when observing a small area of the detector, the isotropic assumption holds. 

\subsubsection{Analytic simulation of the melt pool}
\label{sususec:thermal}
\begin{figure}[h]
	\begin{center}
	\includegraphics[width=.75\textwidth]{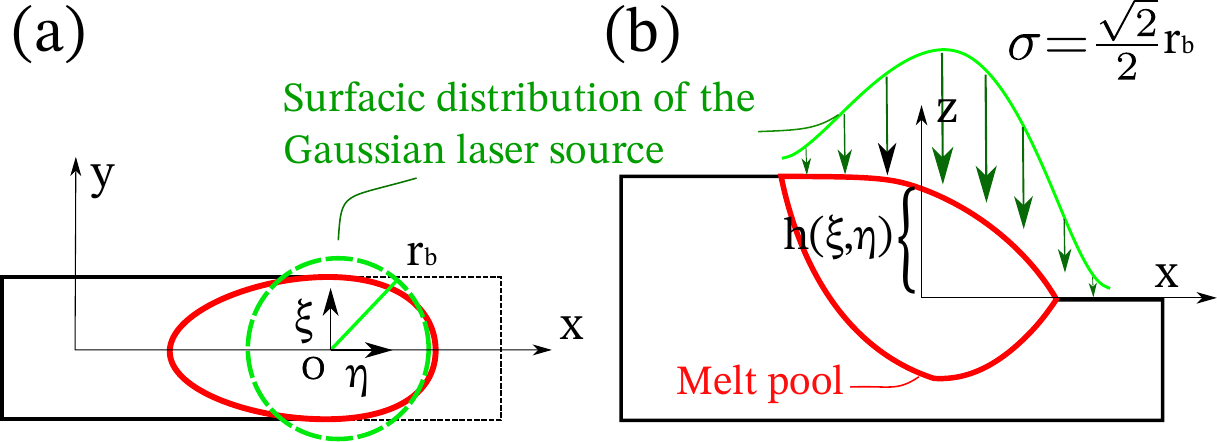}
	\end{center}	\caption{\label{fig:gaussian} (a) Upside and (b) side views of the simplified geometry of the melt pool and the Gaussian laser source applied onto it.}
\end{figure}

The thermal simulation consists in applying a moving Gaussian heat source onto the non-planar surface of the melt pool (figure~\ref{fig:gaussian}). This simulation is inspired by the work of \citet{Fathi2006}.

The geometry of the melt pool is simplified and governed by two equations, one for its height ($h(x,y)$, equation~\ref{eq:height}) and another for its width ($w(x)$, equation~\ref{eq:width}).

	\begin{equation}\label{eq:width}
	\begin{aligned}
	&w(x)|_{z=0} = 
	\begin{cases}
    w_{0}\text{ if } x \leq 0,\\
    w_{0} \biggl(1-\frac{x^{2}}{(w_0/2)^{2})}\biggr)^{\frac{1}{2}} \text{ otherwise},\\
	\end{cases}\\
	&\text{with } x\in[-L2,L1],
	\end{aligned}
	\end{equation}
with :
\begin{itemize}
	\item $w_0$ the maximum width of the melt pool along the $y$ axis (\textit{i.e.} at its center, $x=y=0$),
	\item $L1 + L2$ the total length of the melt pool along the $x$ axis.
\end{itemize}
And
    	\begin{equation}\label{eq:height}
		\begin{aligned}
		&h(x,y) = h(x,0)\left[1-\frac{4y^{2}}{w(x)^{2}}\right],\\
		&x\in[-L2; L1] \text{ and } y\in\left[-\frac{w(x)}{2}, +\frac{w(x)}{2}\right].
		\end{aligned}
	\end{equation}

The Gaussian heat source is applied onto this non-planar surface (figure~\ref{fig:gaussian}) such as :

    \begin{equation}
    		I(r) = I_0 exp\biggl(-\frac{r^2}{\sigma^{2}}\biggr),
    \label{eq:gaussian}
    \end{equation}
    
with : 
    \begin{equation}
    		I_0 = \frac{\beta P_n}{\pi \sigma ^{2} \left[ 1 - exp(-\frac{r_b ^{2}}{\sigma ^{2}}) \right]},
    \label{eq:I0_gaussienne}
    \end{equation}
   
with : 
    \begin{itemize}
	    \item $I_0$ the nominal intensity of the laser (W.m$^{-2}$),
	    \item $r$ the radial distance regarding the center of the laser (m),
	    \item $\sigma$ the effective radius of the laser $\sigma = r_b/\sqrt[2]{2}$ (m),
	    \item $r_b$ the radius of the laser (m),
	    \item $\beta$ the laser's absorptivity,
	    \item $P_n$ the power of the laser (W).
    \end{itemize} 
    
And the integrated laser power $q_0$ (in W) onto the surface can be written in the mobile set of coordinates ($\vec{\xi}, \vec{\eta}$) (equation~\ref{eq:laser}).

\begin{equation}\label{eq:laser}
	q_0 = \int_{A}I(r)dA = \int_{\xi = -r_{b}}^{\xi = r_{b}} \int_{\eta = 0}^{\eta = r_{b}(1-\xi^{2})}I(\sqrt{\xi^{2} + \eta^{2}}) d\xi d\eta,
\end{equation}

Finally, the analytical expression of the temperature then writes : 
	\begin{equation}
		\begin{aligned}
		&T(x,y,z) = \frac{I_0}{2 \pi k} \times
		\int_{\xi = -r_b}^{\xi = r_b} \int_{\eta = 0}^{\eta = r_b(1-\xi^2)}
		\frac{1}{R}exp\biggl(-\frac{\xi^2 + \eta^2}{\sigma^2} \biggr)\\
		&\times exp\biggl( -\frac{v_{laser}(R+(x-\xi))}{2 \alpha} \biggr)
		\, \mathrm{d}\eta \mathrm{d}\xi	,\\
		&R = \sqrt[2]{(x-\xi)^2 + (y-\eta)^2 + (z-h(\xi,\eta))^2}.
		\end{aligned}
	\label{eq:temperature_bain_fondu}
	\end{equation}

with $v_{laser}$ the laser velocity (in mm.s$^{-1}$).

\subsection{From temperature to density}
In order to simulate an X-ray shot of an object, three properties of the given object are required : 
\begin{enumerate}
	\item its geometry,
	\item its chemical composition,
	\item its density.
\end{enumerate}

For this study, the material, SS316L, is homogeneous, which means that its chemical composition is the same (table~\ref{tab:ss316L}). 

\begin{table}[h]
\caption{Composition of SS316L for this simulation. W is the weight percentage of the given element.}\label{tab:ss316L}%
\begin{tabular}{@{}llll@{}}
\toprule
Element & W\%  & Element & W\%\\
\midrule
Fe & 66.3 & Mn & 1.60 \\
Cr & 17.3 & Si & 0.53 \\
Ni & 12.0 & B & 0.02 \\
Mo & 2.23 & C & 0.02 \\
\botrule
\end{tabular}
%\footnotetext{Source: This is an example of table footnote. This is an example of table footnote.}
%\footnotetext[1]{Example for a first table footnote. This is an example of table footnote.}
%\footnotetext[2]{Example for a second table footnote. This is an example of table footnote.}
\end{table}

%    \begin{table}[pos = H]
%      \centering
%      \begin{tabular}[x]{|m{3em}|m{3em}||m{3em}|m{3em}|}
%      \hline
%      \textbf{Element} & W\% & \textbf{Element} & W\%\\
%      \hline
%       Fe & 66.3 & Mn & 1.60 \\
%      \hline
%       Cr & 17.3 & Si & 0.53 \\
%      \hline
%       Ni & 12.0 & B & 0.02 \\
%      \hline
%       Mo & 2.23 & C & 0.02 \\
%      \hline
%      \end{tabular}
%      \caption{Composition of SS316L for this simulation. W is the weight percentage of the given element.}
%      \label{tab:ss316L}
%    \end{table}
    
\begin{figure}[h]
	\begin{center}
	\includegraphics[width=.75\textwidth]{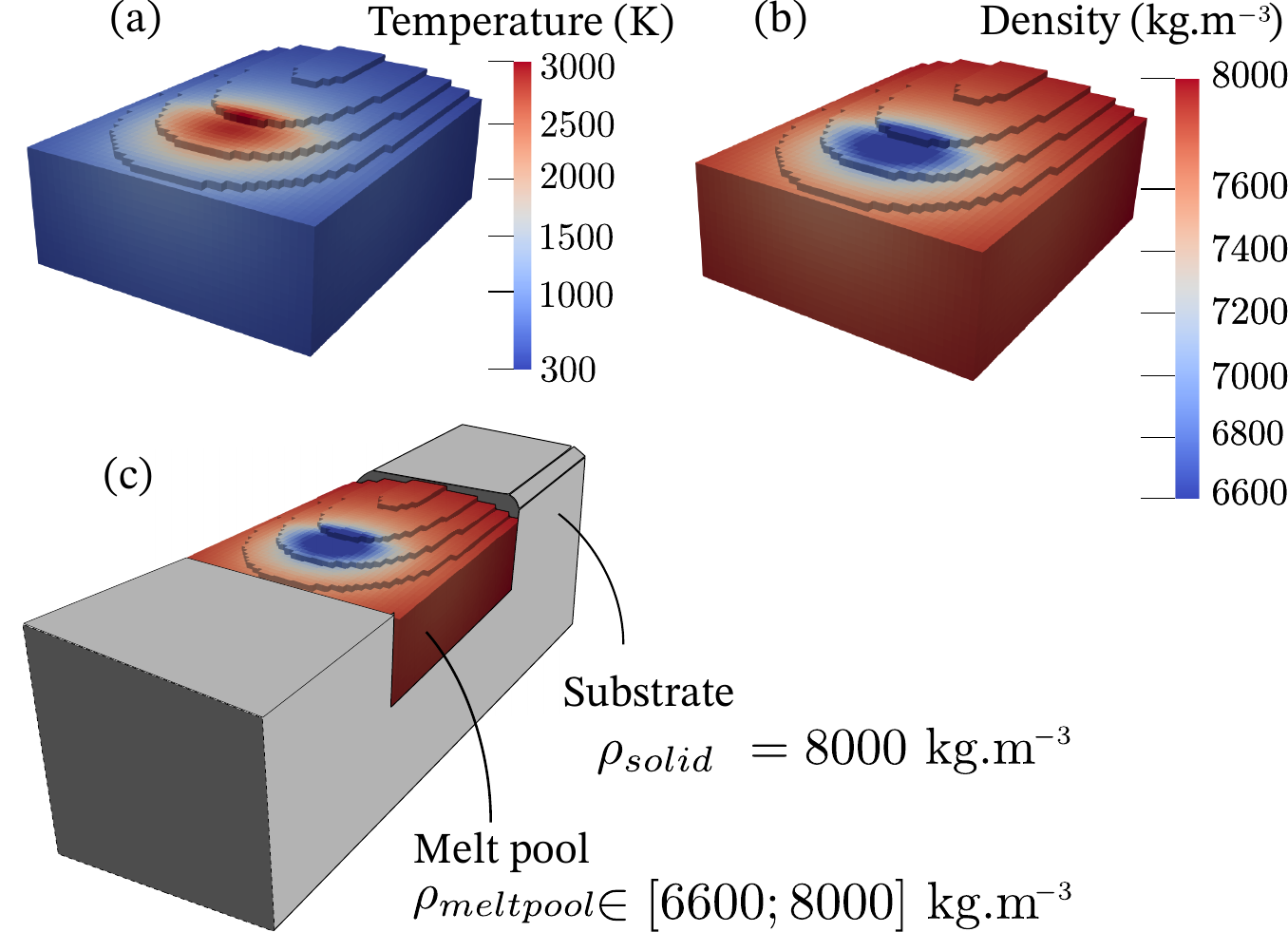}
	\end{center}	\caption{\label{fig:sim_density} Discretization of the melt pool in terms of temperature (a), and density (b). (c) Melt pool within the substrate with a constant density.}
\end{figure}

However, the temperature (\textit{i.e.} the density) of the melt pool varies a lot. In order to simulate an X-ray shot onto it, it has to be discretized into multiple voxels (here, the chosen dimensions were 25$\times$20$\times$10 voxels of size 0.1$\times$0.1$\times$0.1 mm) with each the same chemical composition but a different density. Thanks to the results of the analytical simulation (section~\ref{sususec:thermal}), it is possible to compute the density of each single voxel depending on its temperature (figure~\ref{fig:sim_density}~(a) towards (b)). 
The rest of the substrate, which is solid, has a constant density ($\rho = 8000 kg.m^{-3}$)

All these elements altogether compose the object studied in the X-ray simulation (figure~\ref{fig:sim_density}~(c)). The minimum size of voxels is limited by two factors : 
\begin{itemize}
	\item the computational time. The more single objects, the more computational ressources are needed.
	\item The photons-voxels interactions. The software uses a nonrandom process to simulate the X-ray beam, and the interaction between photons and voxel edges creates computational difficulties.
\end{itemize}

These objects are then used in the X-ray software introduced thereafter.

\subsection{Simulation of the radioscopy}
The radioscopy simulation is structured around three principal components: the X-ray source, the detector, and the studied object. All simulations are conducted using the software \textit{Virtual X-ray Imaging} (VXI) \cite{Duvauchelle2000}. 

\begin{figure}[h]
	\begin{center}
	\includegraphics[width=.75\textwidth]{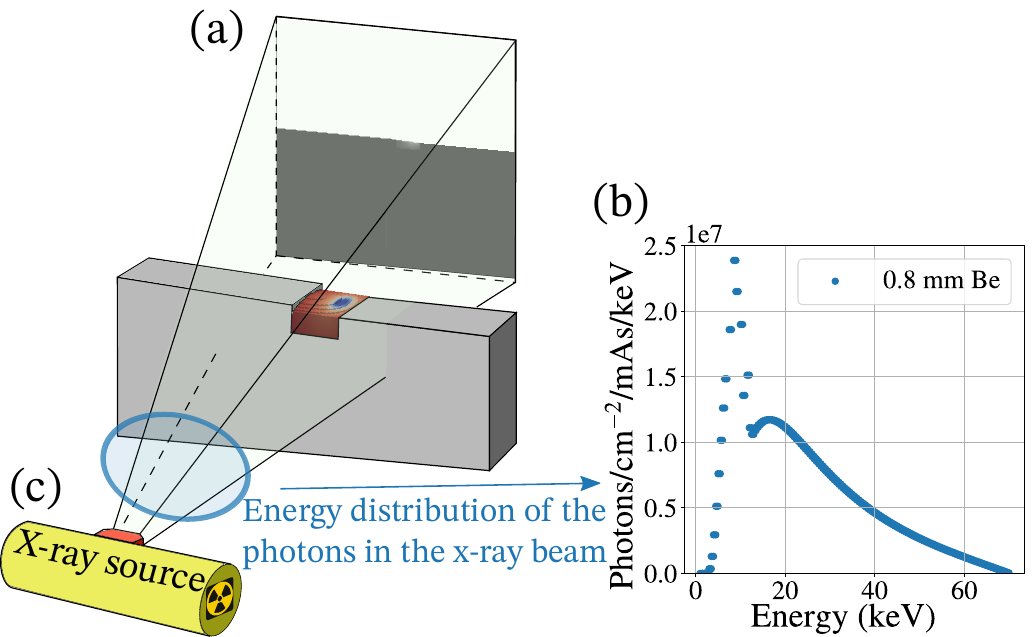}
	\end{center}	\caption{\label{fig:sim_xray} Scheme of a scene on VXI : (a) X-ray detector, (b) X-ray spectrum, (c) X-ray source.}
\end{figure}

The specifications for the X-ray source and detector are derived directly from the experimental setup (section~\ref{sec:source}). The X-ray spectrum (figure~\ref{fig:sim_xray}~(b)) is generated with the Python library \textit{Spekpy} \cite{Poludniowski2021} with a 0.8 mm Beryllium filtration, a voltage of 70 kV, and the target material in tungsten.\\
The detector (figure~\ref{fig:sim_xray}~(a)) is in Gadolinium oxysulfide (Gadox), with 260$\times$260 pixels of size 143$\times$143 $\mu$m. The X-ray beam is isotropic so that it illuminates the entire detector area. 
The spatial arrangement mirrors the experimental conditions with the following distances:
\begin{enumerate}
	\item source to object: 680 mm,
	\item object to detector: 520 mm.
\end{enumerate}

Since these X-ray simulations are deterministic (meaning that two independent simulations with identical parameters will provide the same results), noise and blur (resulting from the focal spot size) are added during a post-processing step. 

\subsection{Results and image processing}
\begin{figure}[h]
	\begin{center}
	\includegraphics[width=.7\textwidth]{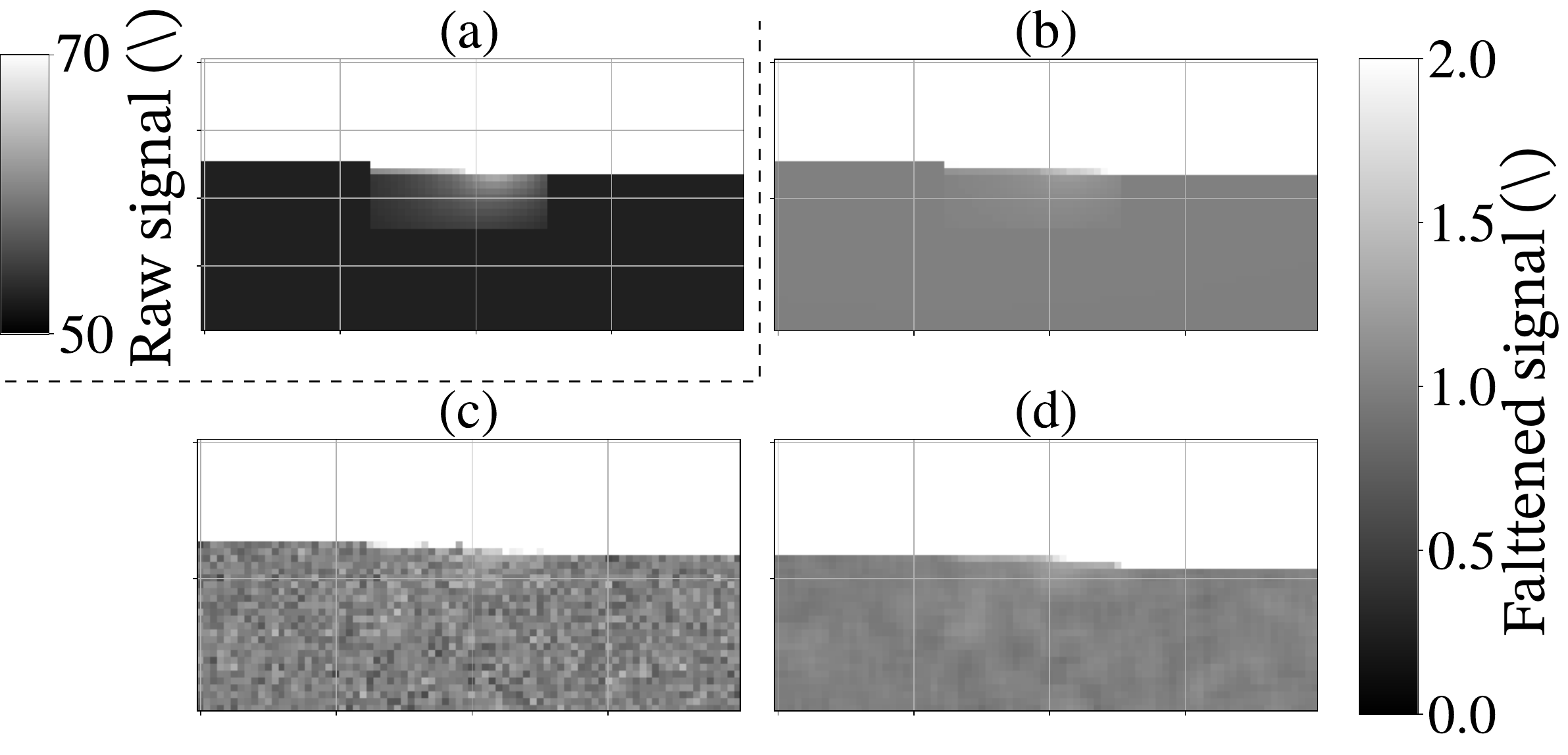}
	\end{center}	\caption{\label{fig:im_process} (a) Original picture. (b) Flattening of the image. (c) Adding noise to the image. (d) Blurred image.}
\end{figure}

The results of these simulations yield 2D images with gray level values representing the energy deposited in the detector due to the quantity of X-rays that pass through the object.
To facilitate a meaningful comparison between these simulations and actual radioscopies, three image processing operations are applied successively:
\begin{enumerate}
	\item the first step involves flattening the image, which consists in dividing the original image by a radioscopy taken without an object (figure~\ref{fig:im_process}~(a) to (b)). This operation normalizes the picture by ensuring that the signal in the air regions is set to 1. Since experimentally the detector saturates at 70 kV and 3 mA, the flattening is accomplished by dividing the original picture by a radioscopy obtained with a 2 mm thick plate of SS316L. As a result, the flattened signal in solid areas equals 1.
	\item The second step involves introducing noise into the radioscopy (figure~\ref{fig:im_process}~(c)). The noise follows a Poisson distribution denoted as $P(\lambda = 6)$. The choice of the $\lambda$ parameter is critical and has been established based on the specific characteristics of the experimental setup, particularly for the source and detector used. 
	\item The final processing step involves blurring the radioscopy (figure~\ref{fig:im_process}~(d)). Given that X-ray sources are not point-like, a natural blur occurs in radioscopies based on the relative positions of the object, source, and detector. The blur is achieved by convolving the image with a matrix of size $k_{blur}\times k_{blur}$ where all entries equal 1. A larger $k_{blur}$ results in a blurrier image. In this setup, $k_{blur} = 5$ is chosen, ensuring that both the simulation and the experimental blurs correspond to approximately 6 pixels.
\end{enumerate}

At this stage, the simulation of a radioscopy of the melt pool is complete. By varying the power and the velocity of the laser during the thermal simulation, it becomes possible to change the size of the melt pool and the contrast variation in the final image. 




\end{appendices}

\bibliography{allbib}% common bib file

\end{document}